%% file: main.tex
\documentclass[journal]{vgtc}                
\input{header}

\usepackage[normalem]{ulem}
\definecolor{mred}{rgb}{.80,.12,.30}
\definecolor{grey}{rgb}{0.5,0.5,0.5}
\definecolor{Purple}{rgb}{.75,0,.85}
\definecolor{pistachio}{rgb}{0.58, 0.77, 0.45}

\newif\ifnotes
\notestrue

\newcommand{\cmt}[1]{\ignorespaces}
\onlineid{331}

\title{At a Glance: Pixel Approximate Entropy\\as a Measure of Line Chart Complexity}

\author{Gabriel Ryan, Abigail Mosca, Remco Chang, Eugene Wu}
\authorfooter{
\item
Gabriel Ryan and Eugene Wu are with Columbia University. E-mail: gr2547@columbia.edu, ewu@cs.columbia.edu. 
\item 
Abigail Mosca and Remco Chang are with Tufts University. E-mail: abigail.mosca@tufts.edu, remco@cs.tufts.edu. 
}

\abstract{
  \input{sections/abstract.tex}

} 


\teaser{
 \centering
 \includegraphics[width=16cm]{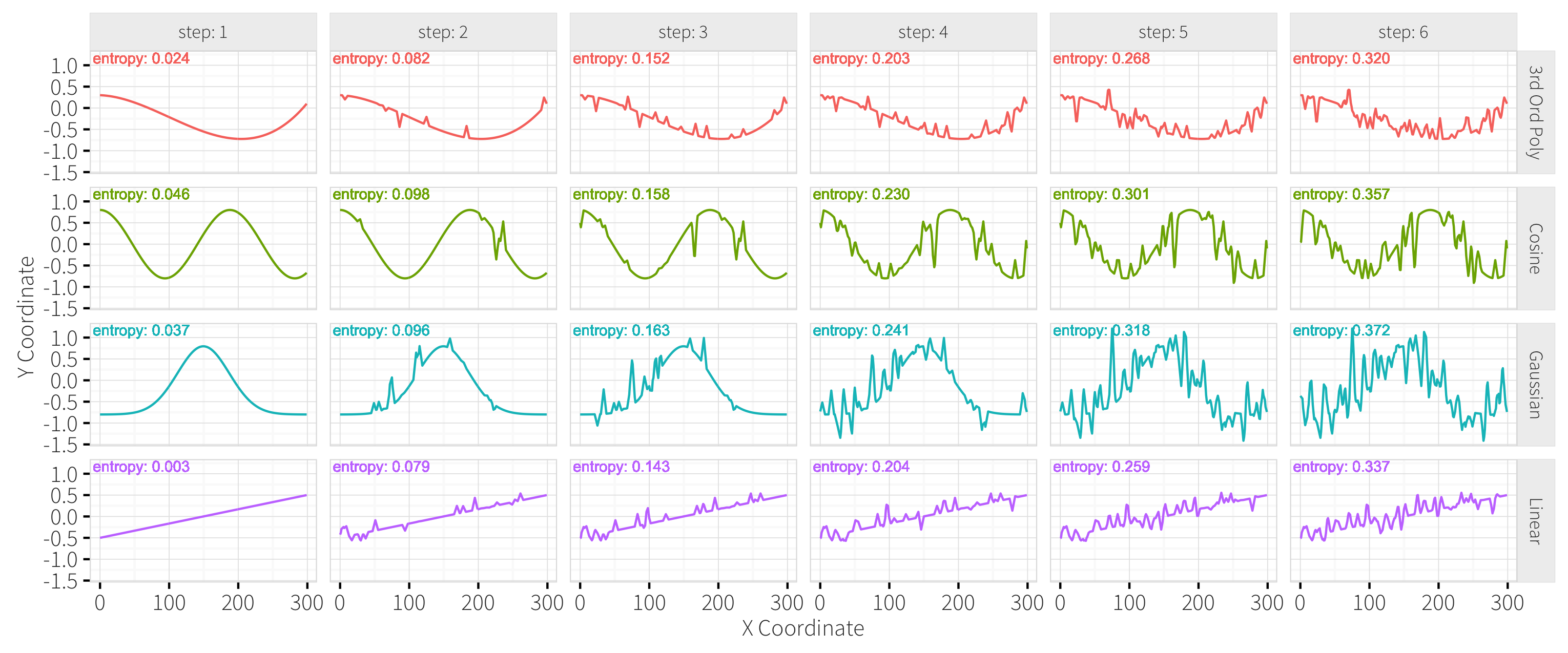}
 \caption{
   Visualization of four line charts: 3rd order polynomial, cosine,  gaussian, and linear (top to bottom).
   Noise is incrementally increased to each line chart from left to right in intervals of proposed approximate entropy measure (shown in upper left of each chart).  
   This paper demonstrates that approximate entropy can be used as a metric for the perceptual complexity of line charts.
 }
 \label{fig:ex_gallery}
 }

\begin{document}


\firstsection{Introduction}

\maketitle


\input{sections/intro.tex}

\input{sections/related.tex}

\input{sections/entropy.tex}

\input{sections/exp.tex}

\input{sections/discussions.tex}

\input{sections/conclusions.tex}


\bibliographystyle{abbrv}
\bibliography{main}

\end{document}

%% file: header.tex
\usepackage{mathptmx}
\usepackage{graphicx}
\usepackage{times}
\usepackage{enumitem}
\usepackage{amsmath}
\usepackage{subcaption}
\usepackage{color}
\usepackage[usenames,dvipsnames]{xcolor}

\usepackage[bookmarks,backref=true,linkcolor=black]{hyperref} 
\hypersetup{
  pdfauthor = {},
  pdftitle = {},
  pdfsubject = {},
  pdfkeywords = {},
  colorlinks=true,
  linkcolor= black,
  citecolor= black,
  pageanchor=true,
  urlcolor = black,
  plainpages = false,
  linktocpage
}

\usepackage[usenames,dvipsnames]{xcolor}
\definecolor{Purple}{rgb}{.75,0,.85}

\newif\ifnotes
\notestrue

\vgtccategory{Application/Design study}

\vgtcinsertpkg

\definecolor{light-gray}{gray}{0.95}
\definecolor{mid-gray}{gray}{0.85}
\definecolor{darkred}{rgb}{0.7,0.25,0.25}
\definecolor{darkgreen}{rgb}{0.15,0.55,0.15}
\definecolor{darkblue}{rgb}{0.1,0.1,0.5}
\definecolor{blue}{rgb}{0.19,0.58,1}

\newcommand{\stitle}[1]{\smallskip\noindent\textbf{#1}}

%% file: sections/abstract.tex
When inspecting information visualizations under time critical settings, such as emergency response  or monitoring the heart rate in a surgery room, the user only has a small amount of time to view the visualization ``at a glance''.  In these settings, it is important to provide a quantitative measure of the visualization to understand whether or not the visualization is too ``complex'' to accurately judge at a glance.
This paper proposes Pixel Approximate Entropy (PAE), which adapts the approximate entropy statistical measure commonly used to quantify regularity and unpredictability in time-series data, as a measure of visual complexity for line charts.  We show that PAE is correlated with user-perceived chart complexity, and that increased chart PAE correlates with reduced judgement accuracy.  `We also find that the correlation between PAE values and participants' judgment increases when the user has less time to examine the line charts.

%% file: sections/intro.tex
Information visualization has traditionally approached research from two directions: how to generate visualization forms, and how those forms are perceived by an end user. The latter branch of research has focused primarily on low-level perceptual questions, such as how a single data point is read and to what degree of accuracy. While this work has been useful and has supported an improved understanding of visualization theory, there remains a missing link between this very bottom-up approach to perception and the top-down approach to generation of visualization forms. In order to connect the two sides of visualization research, a higher level approach to the problem of perception is needed. 

In almost all cases, especially visualization of larger, more complicated data, the first impression of a visualization centers on its overall shape and trend. 
Research in psychology has shown that during the initial glance of a visual stimulus, people perceive the higher-level spatial and functional components of the natural scene \cite{greene2009recognition, greene2006scenes}. Extending this notion, we posit that a similar mechanism applies to the perception of visualizations.
Many real world settings---e.g., emergencies or viewing quickly changing stock prices---are time critical and rely on the user to make judgements based on glances at a visualization.

Although it is anecdotally clear that more complex or noisy visualizations are more challenging to perceive \cite{peng2004clutterRed}, it is unclear how to measure this complexity outside of performing comprehensive user studies.  Developing a measure of visualization complexity that is coherent with perceived complexity can have impact on a number of visualization domains. For example, designers can use the complexity measure to design more effective visualizations, especially for time-sensitive decision making tasks. Similarly,  visualization recommendation engines such as the ``Show Me'' feature in Tableau \cite{mackinlay2007show} can take advantage of the complexity measure to detect when a visualization may be too complex and suggest an alternate design.  To this end, there are a number of desirable characteristics of visual complexity measure:
\begin{enumerate}[leftmargin=*, topsep=0mm, itemsep=-1mm] 
  \item {\it Correlated with perceived complexity: } The measure should correlate with user perception of chart complexity. 
  \item {\it Correlated with noise-levels:} The measure should correspondingly increase when more noise is introduced, since it is computed analytically (rather than through human measurements).   
  \item {\it Predictive of perceptual accuracy:} The measure should correlate with the user's ability to accurately discern patterns in the chart.
  \item {\it Simple:}  The measure should be a single understandable value applicable to arbitrary 1D line charts.
  \item {\it Widely Applicable:} The measure should exhibit the above characteristics across many types of line charts without the need for specialized tuning.
\end{enumerate}

\noindent In this paper, we examine how people perceive, at a glance, the complexity of a line chart visualization, where we consider "at a glance" to be 200ms or less, approximately the amount of time required to visually process a scene ~\cite{rayner2009}. We explore the use of approximate entropy~\cite{pincus1991approximate,pincus1995approximate2}, a statistical measure used to quantify the amount of regularity and unpredictability in time-series data, as a measure to quantify perceptual complexity. 

To this end, we conducted experiments to answer several research questions.  Is approximate entropy an effective measure of time-series visualization complexity? Do users perceive higher approximate entropy visualizations as more complex and lower approximate entropy visualizations as simpler? As the approximate entropy of a visualization increases, does user accuracy in performing visual comparison tasks decrease? When users are given less time to study charts in order to complete a simple identification task, does the approximate entropy measure become more correlated with user accuracy?

We first ran an analytical experiment to compare approximate entropy with synthetically generated noise levels in visualizations.  We then ran four perceptual experiments on Amazon Mechanical Turk~\cite{mturk}: the first asks users to select the most or least complex visualization from a line-up of visualizations with different approximate entropy measures. The result of the study confirms that users perceive visualizations with higher (lower) approximate entropy as more (less) complex.  
The second measures how the visual complexity (as defined by our entropy measure) of a chart affects the user's ability to detect changes in the chart.
The third experiment is similar to the second, but studies the user's ability to identify basic shapes in charts of varying complexity.
We find that visual complexity has a significant and large effect on judgement accuracy for both tasks, and that there is a threshold beyond which judgement accuracy degrades to random chance.  
The last experiment measures the interaction between the amount of time the user has to view a chart (the glance time) with their ability to perform the change detection task in from the second experiment.  As the glance time is reduced to $<200$ms, chart complexity becomes more highly correlated with judgement accuracy.

Our first perceptual experiment provides the basis for using approximate entropy to measure perceived complexity, and the subsequent perceptual experiments are based on real-world use cases. Understanding how people perceive visualization of complex data within a brief period of time can impact real-world usage in multiple ways. In disaster response scenarios, relief workers have limited time to examine data, and need to quickly get a gist of the available information. Similarly, in many real-time health care monitoring tasks, the typical visualization assumption of time to examine data in detail does not hold true. In all of these situations, the data seen at a glance is the only data the user sees. Understanding how this data is perceived is vital for design and evaluation of such cases, and we describe possible applications of this measure in the Discussion section.

%% file: sections/related.tex
\section{Related Work}

The use of Approximate Entropy as a measure of perceptual complexity is related to  research in psychophysics, perceptual psychology, and information visualization (infovis).

\stitle{At a Glance Perception: }
The idea of at a glance perception has been widely studied in perceptual psychology. A well studied area is how people can recognize natural scenes in a short amount of time. Tasks such as rapid scene categorization and object recognition have been found to rely on a broad focus. Greene and Oliva found that in rapid scene categorization, people interpret a scene based on global, ecological properties that describe its spatial and functional aspects rather than by breaking it into objects~\cite{greene2009recognition, greene2006scenes}. Similarly, Biederman et al.\ found that a person's ability to detect an object in a scene is dependent on specific relations between the object and scene as opposed to specific characteristics of the object itself~\cite{biederman1982obj}.  In contrast, data visualization tasks involve recognizing and decoding visually encoded trends and data values.  This paper can be viewed as an initial extension of these ideas towards a potential measure of visual complexity for viewing data visualizations at a glance.

\stitle{Perception of Salient Features: } 
One method for quantifying perception of an image or visualization could be with a method capable of identifying its most salient features, or a measure quantifying the busy-ness of the image. Rosenholtz's work on ``visual clutter", for example, seeks to quantify the amount of clutter in natural image displays. Notably, one of the measures from this work, Subband Entropy, uses entropy to quantify the redundancy in a natural image display and was found to be a reasonable measure of visual clutter~\cite{rosenholtz2007measuring}. Measuring clutter is related to measuring ``visual complexity'' in natural images~\cite{olivia2004identifying}, where a pattern is described as complex if the parts are difficult to identify or separate from each other.

In the visualization community, research has sought to quantify the salient features of a visualization. Scagnostics is one such example by Wilkinson et al., where multiple metrics were proposed to categorize the perception of scatterplots~\cite{wilkinson2005graph,  pandey2016towards}. This work has led to a number of advances, including research that extend the concept of Scagnostics to parallel coordinates~\cite{dasgupta2010pargnostics}, pixel-based displays~\cite{schneidewind2006pixnostics}, and generalized techniques for dimension reduction of high-dimensional data~\cite{bertini2011quality}. However, scagnostics primarily focuses on identifying meaningful relations to visualize with scatter plots, while pixel-based diagnostics focus on intensity-based visualizations like Jigsaw Maps and Pixel Bar Charts. Our work is similar in spirit to these prior work, but instead we focus on the perception of 1D line charts and the quantification of the visual complexity of these visualizations.

Other work has been done on visual analysis and simplification of time series data. Heer et al.\ measure the effect of chart size and layering on speed and accuracy in visual comparison tasks~\cite{heer2009sizing}. ASAP uses kurtosis to guide time-series smoothing to preserve trends and anomalies while reducing cyclic patterns and noise~\cite{rong2017asap}. Numerous measures, such as {\it L1 Local Orientation Resolution}, have been developed to select aspect ratios for line charts~\cite{wang2017aspectratio}. Our complexity measure is complimentary to these approaches and can help guide selection of visualization parameters or smoothing.

\stitle{Perception of Visual Marks and Visual Forms: }
The study of at a glance perception has led to much cross pollination between infovis and perceptual psychology. Substantial work from perceptual psychology suggests that modeling a holistic shape envelope could capture a fundamental aspect of perceptual encodings of visualizations. At the first glance of an image, users tend to focus on the ``big picture''~\cite{KimchiEtc}, which should produce a far more compact representation of the most salient information in the image~\cite{FranconeriOxford}. For line charts, this initial big picture is likely to be the holistic shape envelope that surrounds the values. 
  
Infovis research has studied visualization perception for short glance times. However, the emphasis has been on pre-attentive processing of properties of visual marks~\cite{healey1999large} (e.g., color, orientation, etc).  
Fewer works focus on perception of higher-level visualization forms. Szafir et al.\ recently measured how people can quickly perceive summary statistics (for instance centroid, or density) from visualizations such as scatterplots~\cite{szafir2016four}. They investigated four visual statistical tasks: identification of sets of values, summation across values, segmentation of collections, and estimation of structure. 
Related works show that some visual tasks---such as correlation from scatterplots~\cite{rensink2011rapid}, mean size of homogenous elements in an array~\cite{chong2003means}---occur within the pre-attentive processing phase ($<$200ms).

Another related area is the interplay of glance time and a person's processing of visual information. Based on findings from cognitive science, it may be that shorter glance times are difficult for users because they need to make use of short-term memory to perform visual analysis~\cite{kosslyn1989understanding}.  Short-term memory is limited~\cite{cowan2010magical}, decays over time~\cite{brown1958some}, and is expensive to use. For instance, in an experiment conducted by Ballard et al., subjects \emph{serialized} their tasks in order to avoid using short-term memory~\cite{ballard1995memory}.  This relationship between latency and memory is consistent with other recent papers in the visualization community related to the measure of memorability~\cite{lipford2010helping, borkin2013makes}.  We study the interaction effects between glance time and the user's ability to perform basic visual judgements.

\stitle{Use of Entropy in Visualization: }
There has been a long history of the use of entropy in the field of visualization. A recent book by Chen et al.\ summarizes a wide range of applications of entropy and information theory in visualization~\cite{chen2016information}.
Further, scientific visualization research has leveraged entropy to e.g., allocate computational resources~\cite{wang2008importance}, choose rendering properties~\cite{wang2011information}, and select compression techniques~\cite{tavakoli1993imageComp}.

In the field of information visualization, similar to scientific visualization, entropy has been used to measure the amount of noise and information in the data. For example, Chen and Jaenicke use entropy to detect the amount of information in a visualization, and ways to optimize the design of visualization~\cite{chen2010information}. Dasgupta et al.\ use entropy to measure visual uncertainty to preserve privacy in parallel coordinates~\cite{dasgupta2013measuring}. Biswas et al.\ use entropy to measure importance and model relationships of variables in a multivariate dataset~\cite{biswas2013multi}. Lastly, Karloff and Shirley use entropy to determine an optimal summary tree for large node-weighted rooted trees~\cite{karloff2013summTree}. 

However, these works primarily use entropy to measure data quality, rather than as a proxy for visualization {\it perception}.
In fact, Chen et al.\ note that while related, information theory is about efficient communication rather than perceptual and cognitive processes~\cite{chen2016information}, and that compressing the data in a visualization into the minimum set of bits may not result in the best visual representation.
In contrast, this paper establishes, quantifies, and explores the relationship between information theoretic notions of randomness (approximate entropy) with graphical perception.

Recent work by Rensink et al. theorized that the perception of correlation in scatterplots can be explained by measuring the entropy of the data points in the scatterplot~\cite{rensink2017}. Although our study of approximate entropy focuses on the perception of complexity of line charts, we share the same research goal of evaluating entropy as a possible perceptual complexity measure.

%% file: sections/entropy.tex
\section{A Line-chart Complexity Measure}

\begin{figure}[bt]
  \centering
  \includegraphics[width=.85\columnwidth]{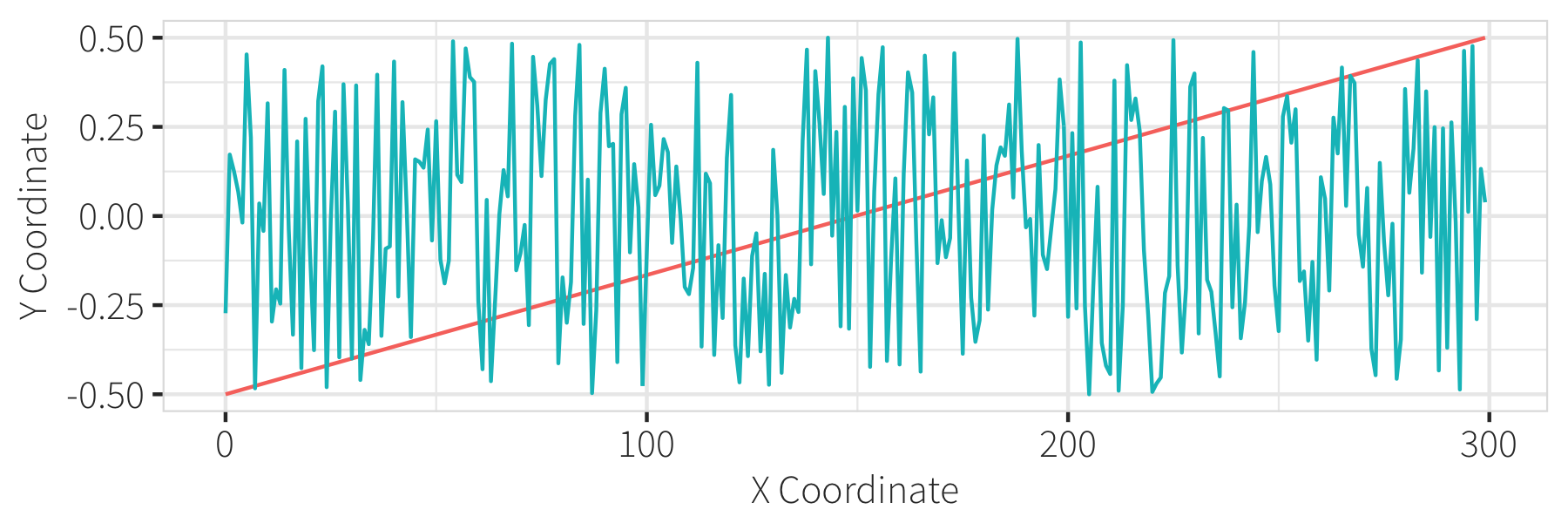}
  \caption{Linear line and its shuffled variant will both have the same level of entropy based on Shannon Entropy.  However the linear line is intuitively less complex than its shuffled variant. 
  }
  \label{fig:ex_shuffled}
 \end{figure}

 Since entropy has traditionally been used to measure the amount of ``disorder'' in data and has been utilized in perceptual psychology as a proxy of ``visual clutter'', it is reasonable to apply the same concept to measure the complexity of a line-chart visualization. However, the classic measure of entropy is computed over an unordered set of values, and does not work for ordered visualizations. Figure~\ref{fig:ex_shuffled} illustrates an example---both the linear line (red) and its shuffled variant (blue) have the exact same entropy measure.  
 
 We therefore surveyed the signal processing, information theory, and statistical modeling literature and identified 8 candidate entropy measures for line charts.  We then selected a subset based on the five desirable characteristics of a complexity measure described in the Introduction.  These were selected to described their scope and ease of application, as well as relation to user perception, rather than reliant on any particular notion of complexity.  For instance, we do not want to assume that ``more jagged'' shapes are more complex.   
 
In reviewing possible measures, we encountered a wide variety of complexity measures with correspondingly varying definitions of complexity for a given chart. Our goal when conducting this review was not to attempt to define complexity itself, but to identify a measure that approximates how well users will be able to perceive and use a chart. To this end, we only imposed two analytical constraints based on the first and fourth desired criteria: that the measure should identify more noisy signals as more complex, since noisy signals are anecdotally more difficult to perceive~\cite{peng2004clutterRed}, and that the measure computes a single scalar number.

\begin{itemize}[leftmargin=*, topsep=0mm, itemsep=-1mm] 
  \item {\it Signal to Noise Ratio}: SNR is a signal processing method that measures the relative power of the desired signal to the noise overlaying that signal. 

  \item {\it Auto Correlation}: A method commonly used in Signal Processing that quantifies the amount of repetitions in a time series by measuring the correlation of the time series with a lagged version. Since a more random time series will have less repetition, this approximates the amount of randomness in the time series.

  \item {\it Fourier Analysis}: Fourier Analysis~\cite{bracewell1986fourier} is also a common method in signal processing. It transforms the data into the frequency domain, making it possible to quantify the extent to which the data is made up of different frequencies. Since more random data should result in higher frequency changes, this can also be used to approximate randomness by measuring the high frequency components of the signal.

  \item {\it Approximate Entropy}: Approximate Entropy~\cite{pincus1995approximate2,pincus1991approximate,ho1997predicting} is a robust statistical measure of repetitiveness. It is based on randomness statistics for chaotic functions from Information Theory. Like Auto Correlation, it measures how much components of the signal repeat, but derives an entropy measure for time series data.

  \item {\it Sample Entropy and related statistics}: These modifications of Approximate Entropy remove possible biases~\cite{RichmanH2039,Yentes2013,fuzzyentropy2009}.

  \item {\it Multiscale Entropy}: This applies approximate or sample Entropy multiple scales to analyze how a signal may be more or less chaotic at varying scales~\cite{multiscale_entropy2002}.

  \item {\it Flattened Signal Length}: This method can be understood as 'stretching' the data until it becomes flat, intuitively, more complex charts will result in longer flattened lines~\cite{batista2011complexity}. 

  \item {\it Sequential Modeling}: Measuring the ability of a Hidden Markov Model to predict the signal. This is a classic statistical modeling method for sequential data. Intuitively, a signal that cannot be easily modeled will be more random~\cite{arias2015entropies}.

\end{itemize}

\noindent Although we started with these measures, we found that most were not appropriate, or did not satisfy the desirable criteria listed in the Introduction. Fourier analysis and auto-correlation did not analytically correlate closely with added noise to a given line chart (e.g., those shown in Figure~\ref{fig:ex_gallery}).  Hidden markov models require tuning a number of hyperparameters---such as dimensionality and the initial state---that vary across charts. Similarly, SNR requires a pre-existing model for the desired signal that is unlikely to exist in a real world application.  Flattened Signal Length does not account for repetition in the data, thus it can, for example, assign a sinusoid a higher complexity than random data, although the random data visually appears more complex.

Approximate, Sample, Fuzzy, and Multiscale Entropy are all similar, in that Sample Entropy and Fuzzy Entropy are both bias corrected versions of Approximate Entropy, and Multiscale Entropy applies Approximate/Sample Entropy at multiple scales. Multiscale Entropy violates the goal of a {\it simple} measure because it generates measurements for each scale.   We found that in practice, Approximate Entropy and Sample Entropy tend to be very close in value, however, Sample Entropy is sometimes not defined for low entropy charts. We therefore selected Approximate Entropy as the the candidate measure for study in this paper.

\subsection{Approximate Entropy}

Approximate entropy is a family of system parameters and related statistics developed by Pincus to measure changes in system complexity~\cite{pincus1991approximate}. In particular, the statistic is designed to be effective at distinguishing complexity in low dimensional systems when only relatively few (tens to low thousands) points are available. This property makes it an effective measure for line chart visualizations, which are low dimensional and often contain hundreds of pixels (at most one point per pixel).

\begin{figure}[tb]
  \centering
  \includegraphics[width=.95\columnwidth]{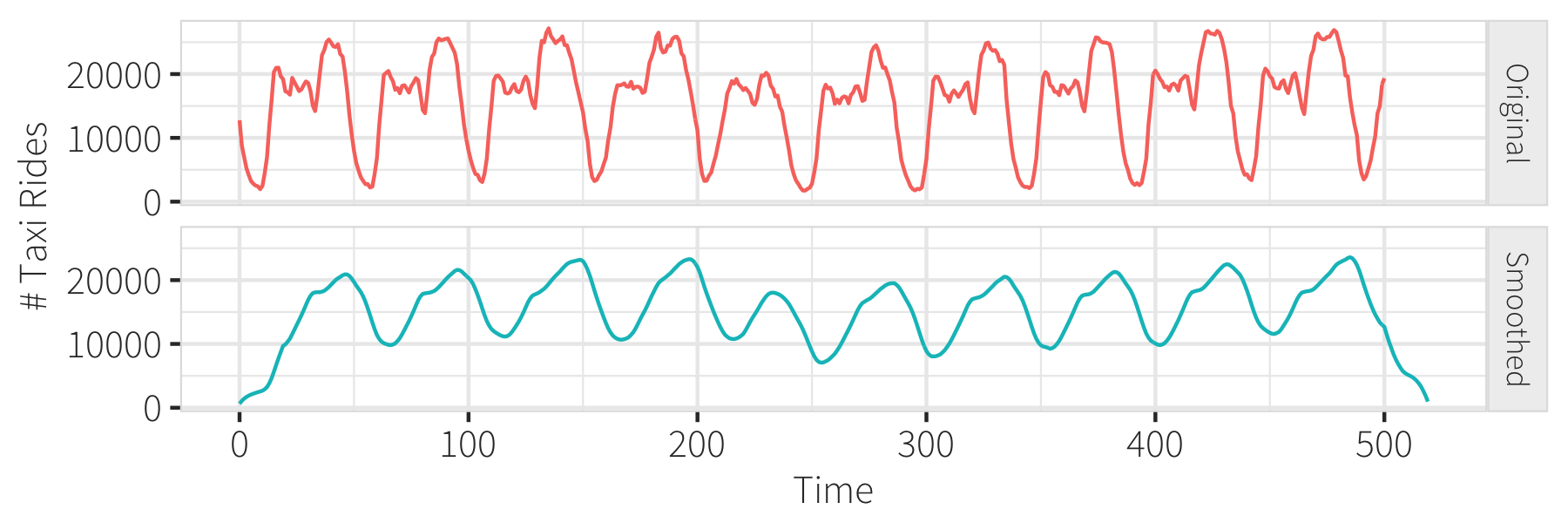}
  \vspace*{-.1in}
  \caption{Example of time series taxi usage in New York City and a smoothed variant. }
  \label{fig:taxi}
 \end{figure}

Approximate entropy quantifies the unpredictability of changes in a series of points. Intuitively, a series of points with more repeated patterns is easier to predict.  Approximate entropy reflects the probability that such similar patterns will not be repeated.  A line chart without any repetitions, such as stochastic noise will have a very high entropy. In contrast, taxi cab demand in New York City (Figure~\ref{fig:taxi}, top), which exhibits very large changes between high and low demand periods, will tend to have a lower entropy because the demand follows a regular daily and weekly pattern.  Notice though that the smaller variations in the original plot cause it to have larger entropy than the smoothed version (Figure~\ref{fig:taxi}, bottom). 

Conceptually approximate entropy is computed using a sliding window approach.  
Given $N$ samples of a continuous curve, each window of size $m$ is compared with every other window of the same size.
If there are many pairs of similar windows, then there is more regularity in the curve and the score should be lower.

\begin{figure}[bt]
  \centering
  \includegraphics[width=.75\columnwidth]{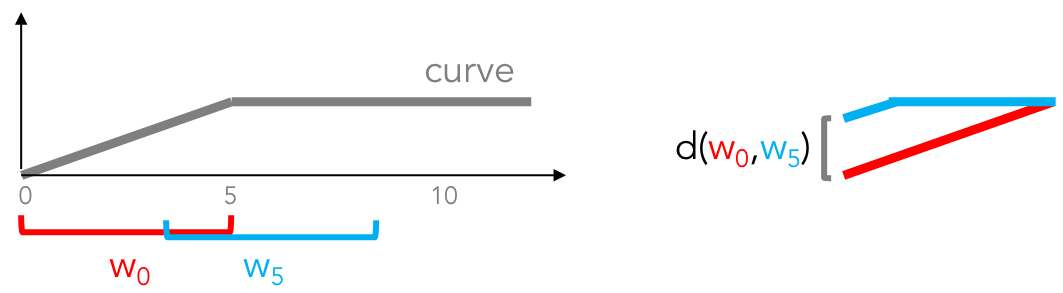}
  \vspace*{-.1in}
  \caption{Example of a curve (grey) and two windows ($w_0$ and $w_5$) on the left.  The right side shows that the distance function $d(w_i, w_j)$ is simply the maximum difference between aligned pairs of y values.}
  \label{fig:approxent}
 \end{figure}

Formally, let $y_i$ be the $i^{th}$ sample of the input curve, and $w_i^m = [y_k | k \in [i, i+m]]$ be the $i^{th}$ window of size $m$.
Thus, there are a total of $W=N-m+1$ possible windows of size $m$ for a line with $N$ values.
Figure~\ref{fig:approxent} illustrates a grey curve and two examples windows $w_0^5$ and $w_5^5$ of size $m=5$.
We define the distance between two windows as the maximum difference between aligned values:
\begin{eqnarray}
  d(w_i^m, w_j^m) = max_{k\in[0,m]} |y_{i+k} - y_{j+k}|
\end{eqnarray}
Then, the similarity $S^m_i(r)$ for a given window $w_i$ is defined by the percentage of windows whose distance is below a threshold $r$\footnote{$r$ is in the same units as $y$.}:
\begin{eqnarray}
  S^m_i(r) = \frac{|\{w_k | k\ne i \wedge d(w_k^m, w_i^m) < r\}|}{W}
\end{eqnarray}
We can now combine the similarity scores for all possible windows into $\Phi^m(r)$ by summing their log transforms. This can be understood as taking the log probability that windows will be closer than $r$. Note that there is a $\Phi$ for each user-defined window size and threshold.
\begin{eqnarray}
  \Phi^m(r) =\frac{1}{W} \sum_{i=1}^{W} log(S^m_i(r))
\end{eqnarray}
Pincus~\cite{pincus1991approximate} originally defined the approximate entropy to be the difference $\Phi^m(r)-\Phi^{m+1}(r)$ as the number of samples from the curve $N$ increases to $\infty$. Intuitively, this difference measures the increased probability that sequences will be greater than $r$ apart when the sequences length increases by one.  
\begin{eqnarray}
\textrm{E}(m, r) = \textrm{lim}_{N \rightarrow \infty} [\Phi^m(r) - \Phi^{m+1}(r)]
\end{eqnarray}
However, sampling infinite points is not realistic, and thus approximate entropy is estimated using a fixed $N$:
\begin{eqnarray}
\textrm{E}(m, r, N) = [\Phi^m(r) - \Phi^{m+1}(r)]
\label{eq:pae}
\end{eqnarray}
This is a natural fit for visualizations, where we can set $N$ to the width of the chart in pixels.

\subsection{Pixel Approximate Entropy}

We define the {\it pixel approximate entropy} as the approximate entropy of a line chart visualization. To do so, we use the following procedure:
\begin{enumerate}[leftmargin=*, topsep=0mm, itemsep=-1mm] 
\item Scale the dataset by mapping its values into the visual domain as positional variables, so that the x and y data values are in terms of pixels.
\item Construct a vector  $Y = [y_i | i \in [0, N]]$ where $y_i$ is the pixel y-coordinate for the curve's $i^{th}$ pixel along the x coordinate.  $N$ is the pixel width of the chart.
\item Compute $E(m, r, |Y|)$. 
\end{enumerate}

\noindent Pixel approximate entropy (PAE) calculates approximate entropy based on pixels in the visualization itself. The benefit is that it is independent of the data complexity, and provides a consistent range of entropy values for a given chart resolution.

Finally, note that we have chosen to compute PAE as a single global complexity measure based on the positionally encoded data in the line chart. Alternative measures that e.g., capture both local and global complexities, or account for additional visual encodings, are promising extensions of this work.

\subsection{Examples of Pixel Approximate Entropy}

To provide a sense of PAE values for different types of line charts, Figure~\ref{fig:ex_gallery} illustrates four base visualized curves and their PAE values (the text in the upper left of each plot shows the PAE measure).  
We show a third order polynomial (top), cosine function (2nd row), Gaussian distribution (3rd row), and a linear line (bottom).  
We chose these curves because they are representative of common visualized data in practice.  
The linear line is the simplest shape that users commonly encounter and we chose it for its simplicity.
The gaussian distribution is arguably the most well recognized distribution, and models natural phenomena such as sizes of living tissue (e.g., length, height, weight), stock distributions~\cite{black1976taxes}, intelligence~\cite{herrnstein2010bell}, and other societal and scientific data.
The third order polynomial represents a more complex pattern that is commonly used in scientific models such as thermodynamics~\cite{smith1975introduction} and kinematics.
Finally, the cosine function represents cyclic patterns such as heart beats, and temperature over time.

Each column in the figure, going from left to right, adds more random noise to the curve making it more ``complex''.  We can see that the curves become seemingly more random, however the overall shapes are still evident.

\begin{figure}[ht]
  \centering
  \begin{subfigure}[t]{.48\columnwidth}
    \centering
    \includegraphics[width=0.5\columnwidth]{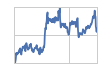}
    \caption{Original Data, Entropy: $0.196$}
  \end{subfigure}
  \begin{subfigure}[t]{.48\columnwidth}
    \centering
    \includegraphics[width=\columnwidth]{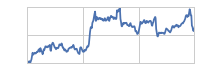}
    \caption{Width $\times 2$, Entropy: $0.0906$}
  \end{subfigure}\\
  \vspace*{0.1in}
  \begin{subfigure}[t]{.48\columnwidth}
    \centering
    \includegraphics[width=0.5\columnwidth]{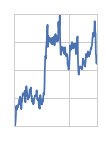}
    \caption{Height $\times 2$, Entropy: $0.459$}
  \end{subfigure}
    \begin{subfigure}[t]{.48\columnwidth}
    \centering
    \includegraphics[width=\columnwidth]{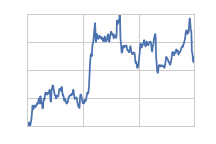}
    \caption{Scale $\times 2$, Entropy: $0.229$}
  \end{subfigure}
  \caption{Scaling effects on Pixel Approximate Entropy. Increasing the height of the chart will increase PAE, while increasing chart width will decrease PAE. }
  \label{fig:scaling_effect_example}
\end{figure}

To demonstrate how Pixel Approximate Entropy works in practice, we provide several examples of its behavior. Figure~\ref{fig:scaling_effect_example} shows how altering the scale of a chart effects its entropy measure. Increasing the relative height of the chart will increase the entropy, while increasing the width will decrease entropy. These scaling effects can be understood as either increasing the relative noisiness of the chart (for height), or having a smoothing effect by stretching the chart out (for width).

\begin{figure}[ht]
  \centering
  \begin{subfigure}[t]{.48\columnwidth}
    \centering
    \includegraphics[width=\columnwidth]{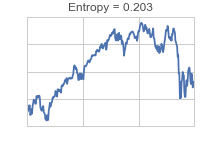}
    \caption{S\&P 500 price data.}
    \label{fig:sp_price_ex}
  \end{subfigure}
  \begin{subfigure}[t]{.48\columnwidth}
    \centering
    \includegraphics[width=\columnwidth]{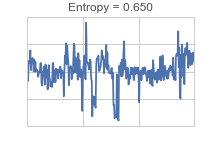}
    \caption{S\&P 500 trade volume data}
    \label{fig:sp_volume_ex}
  \end{subfigure}\\
  \vspace*{0.1in}
  \begin{subfigure}[t]{.48\columnwidth}
    \centering
    \includegraphics[width=\columnwidth]{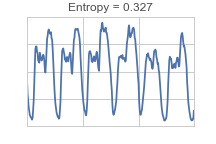}
   \caption{Taxi ride volume data in NYC.}
    \label{fig:taxi_ex}
  \end{subfigure}
    \begin{subfigure}[t]{.48\columnwidth}
    \centering
    \includegraphics[width=\columnwidth]{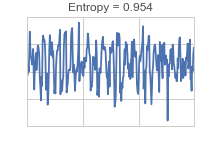}
    \caption{EEG seizure data.}
    \label{fig:eeg_ex}
  \end{subfigure}
  \caption{Examples of Pixel Approximate Entropy applied to real world data. More visually noisy data, such as stock trading volume and EEG data collected during a seizure, has higher approximate PAE.}
  \label{fig:real_data_examples}
\end{figure}

Figure~\ref{fig:real_data_examples} depicts PAE for several real world data sets. In practice PAE can be interpreted as the amount of space on the chart that is taken up by {\it unpredictable} data. Charts of data that exhibit less noise, such as the S\&P 500 stock pricing data (Figure~\ref{fig:sp_price_ex}) and NYC taxi ride volume data (Figure~\ref{fig:taxi_ex},) have relatively low PAE. Charts with more irregular, noisy data, such as the S\&P 500 trade volume data and EEG seizure data in Figures~\ref{fig:sp_volume_ex} and~\ref{fig:eeg_ex}, have much higher PAE.

%% file: sections/exp.tex
 \section{Experiments}

This section presents four experiments to evaluate Pixel Approximate Entropy as a visual complexity measure consistent with the desired criteria in the Introduction. We translated these criteria into the following hypotheses.  Each hypothesis corresponds to one experiment, and we describe the hypotheses in more detail in the corresponding experiment subsection:  

\begin{itemize}[leftmargin=*, topsep=0mm, itemsep=-1mm] 
	\item \textbf{H1:} There is a statistically significant correlation between PAE and the amount of noise added to the chart.
	\item \textbf{H2:} PAE is an effective measure of perceived complexity, such that there is a statistically significant correlation between participants' perception of complexity and the line chart's PAE.
	\item \textbf{H3:} Varying chart PAE affects participants' ability to perform the visual task of identifying changes in a line chart.
	\item \textbf{H4:} Varying chart PAE affects participants' ability to perform the visual task of identifying the base function of a chart.
	\item \textbf{H5:} Reducing the amount of time participants are given to study line charts, and the PAE of a line chart ``at a glance'' affects comparison accuracy on charts.  
\end{itemize}

Experiment 1 verifies that PAE is correlated with the amount of noise added to chart (\textbf{H1}).  We then present four user studies that use both controllable synthetic charts as well as charts from real-world medical and financial datasets to  evaluate the user's ability to perform perceptual tasks at varying PAE levels.  Experiment 2 uses the Line-Up~\cite{wickham2010graphical} protocol to test PAE's correlation with perceived complexity (\textbf{H2}).  Experiment 3 and 4 test the effect of PAE on two visual comparison tasks---matching identical charts and identifying the underlying shape in a chart (\textbf{H3,4}).  Experiment 5 studies the interaction between glance time and PAE when matching identical charts (\textbf{H5}).

\subsection{Experimental Setup Overview}

We now describe the shared experiment setups.

\begin{figure}[tbh]
  \centering
  \includegraphics[width=.55\columnwidth]{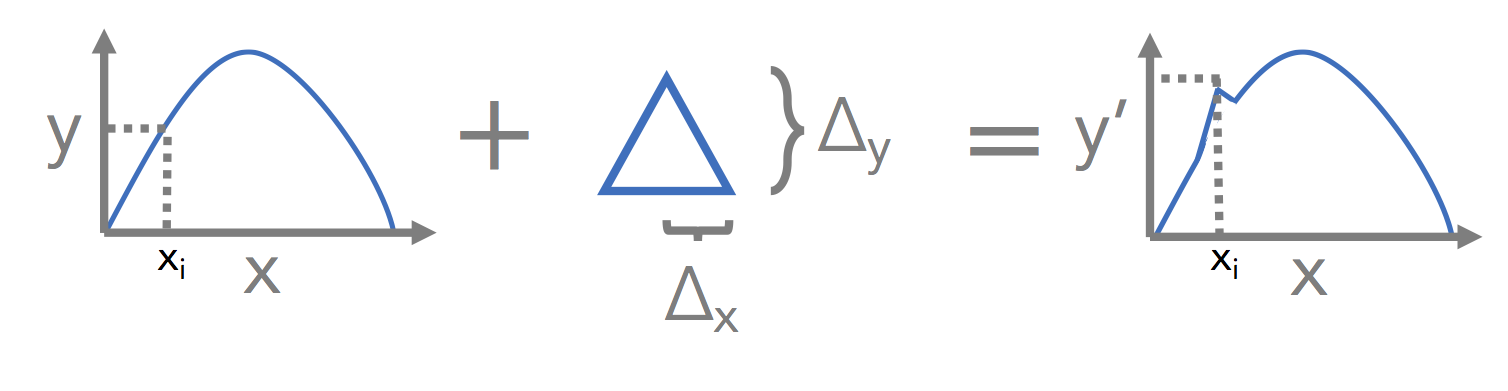}
  \caption{Noise $\Delta_y$ is added at a point along the input curve and interpolated by $\Delta_x$ to each side (e.g., add triangle to a region on curve).}
  \label{fig:addnoise}
\end{figure}

\stitle{Noise Generation:} For the synthetic data used in our experiments, we systematically introduce noise to control the PAE value of a chart.  To do so, we iteratively add noise to a baseline chart until its measured PAE reaches the desired value.  Figure~\ref{fig:addnoise} depicts how triangle noise is added to a given curve. We sample the magnitude of the noise $\Delta_y$ from a uniform distribution $U(-\sigma, \sigma)$, where $\sigma$ is the standard deviation of the data, and add that value to the $y_i$ value of a random x coordinate $x_i$. Thus we set $y' = y + \Delta_y$.  We then linearly interpolate $y'$ with the value of the curve at $x_i\pm\Delta_x$.  

\stitle{Approximate Entropy Parameter Selection:}
Approximate entropy is parameterized by $m$ and $r$ (see Equation~\ref{eq:pae}), which we set to $m=2$ and $r=20$ in our experiments.  We selected these values by synthetically generating a training set of line charts with varying amounts of generated noise and found values that maximized the average PAE-to-noise level correlation across the training set. 

Specifically, we started with four basic curves, added varying amounts of noise (see Figure~\ref{fig:ex_gallery} for examples), and rendered the results at different visualization resolutions ($\{ 100\times 150, 200\times 300, 400\times 600\}$).  We swept the $m$ and $r$ parameters and found that the $m=2, r=20$ parameter settings were most robust across the visualizations. For consistency, all charts are $300\times200$ pixels.

\stitle{User Study Setup:} We used Amazon Mechanical Turk~\cite{mturk} to recruit participants for the user studies. In the default  setup, participants  were given a consent form, a training exercise that introduced the task, a brief qualification test, and a demographics survey at its completion. Participants were paid per task at an estimated rate of $\ge\$8.00$ per hour, and all assignments included a 1.00 USD bonus for completing all tasks. For experiments 3 through 5, which involve simulated visual tasks, participants were given an overall time limit based on the assumption they would spend at most 20 seconds per task, with an additional 3 minutes to read the instructions and complete survey. This time limit was intended to keep participants focused while allowing for the possibility of minimal interruptions. In practice, most participants completed each task in less than 4 seconds.

Due to the simplicity of the task in Experiment 2, the qualification task was not included, and participants were simply paid a bonus based on the number of correct answers. Workers were required to have an approval rating of at least $80\%$ and be residents of the United States. We selected an approval rating less than $95\%$ to engage workers that would attempt to perform the tasks quickly and without too much effort---thereby simulating a routine visual task.

\input{sections/exp1.tex}

\input{sections/exp2.tex}

\input{sections/exp3.tex}

\input{sections/exp4.tex}

\input{sections/exp5.tex}

%% file: sections/exp1.tex
\subsection{Experiment 1: PAE and Synthetic Noise}

\begin{figure}[thb]
  \centering
  \includegraphics[width=\columnwidth]{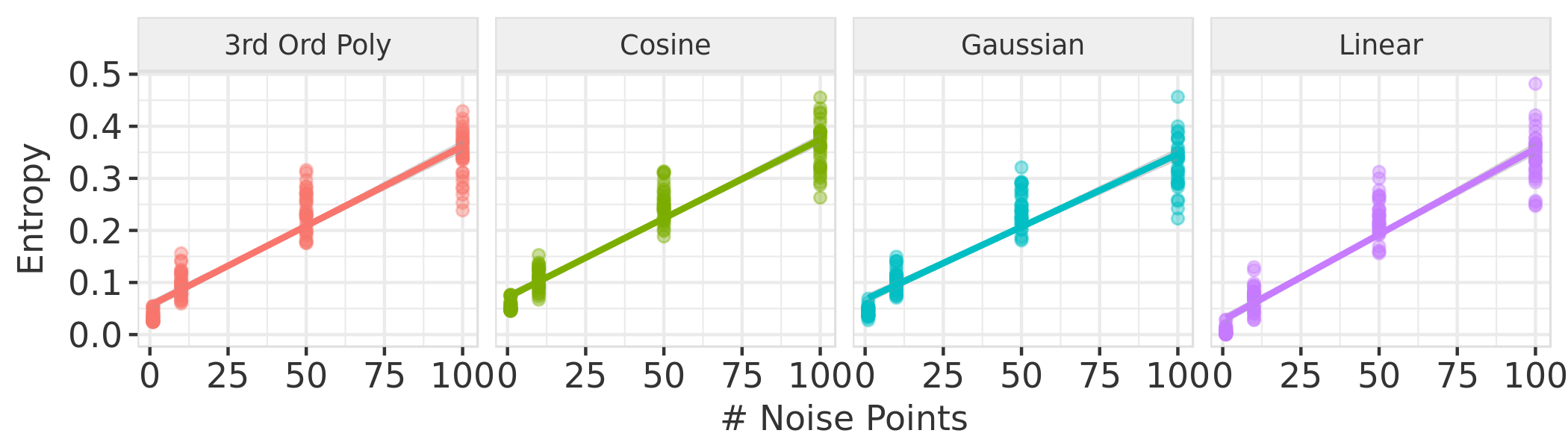}
  \vspace*{-.1in}
  \caption{Correlation between noise level and entropy measure}
  \label{fig:exp1_summary}
 \end{figure}

Our experiments rely on controlled experiments that carefully vary the ``complexity'' of a chart by adding or removing noise as described above.  Since we use PAE as a proxy for complexity, this experiment first establishes the correspondence between the amount of noise added to different basic line chart functions, and the resulting chart's PAE.  We select basic functions that tend to appear in many different types of charts: linear, Gaussian distribution, cosine, and 3rd order polynomial.

Figure~\ref{fig:exp1_summary} plots the relationship between noise and PAE.  Additionally, for each base function, we test correlation through linear regression between the independent (amount of noise) and dependent (PAE) variables.
We evaluate the regression with a t-test against the null hypothesis that there is no correlation, and by the coefficient of determination $R^2$. 
To summarize the t-test and $R^2$ for each base function: 
3rd Order Polynomial ($t=37.3$, $p<.0001$, $R^2=0.91$),
Linear ($t=37.8$, $p<.0001$, $R^2=0.91$),
Cosine ($t=36$, $p<.0001$, $R^2=0.90$),
Gaussian ($t=31.7$, $p<.0001$, $R^2=0.88$).

Based on these findings, we accept  \textbf{H1}, that the PAE of a chart correlates closely with the amount of noise added to the chart. 
We use this result in the synthetic data generation used in other experiments, in which we add noise to control the PAE of a given visualization.

%% file: sections/exp2.tex
\subsection{Experiment 2: PAE and Perceived Vis Complexity}

Does PAE correlate with user perceived notions of visualization complexity? In this experiment, we test the hypothesis that PAE is positively correlated with perceived line chart complexity. To understand whether PAE is effective beyond synthetically added noise, we run two studies---one using synthetic charts and one using charts drawn from medical and stock datasets.  
We hypothesize that participants' identification of the most and least complex chart from a line up will correlate with \textbf{(H2.1)} the chart's PAE, and \textbf{(H2.2)} the underlying base function of the chart. 

\stitle{User Tasks:} 
We use two user tasks: one for charts that vary PAE using synthetically generated noise, and one for charts of real data that naturally have varying PAE levels.  For both, the training page informed the participant that she would be shown 20 (for synthetic noise, or 16 for real data) sets of similar charts, and asked to pick the most or least complex chart in a given set. It also showed a figure of three example charts and labeled the most and least complex. Participants then completed the 20 (or 16) judgments. Each judgment consisted of selecting the most or least complex chart out of a set of eight charts with varying PAE. Because we are interested in discovering whether or not a correlation exists between perceived complexity and PAE we use the LineUp method from~\cite{wickham2010graphical} for this experiment. 

\begin{figure}[htb]
  \centering
  \includegraphics[width=.8\columnwidth]{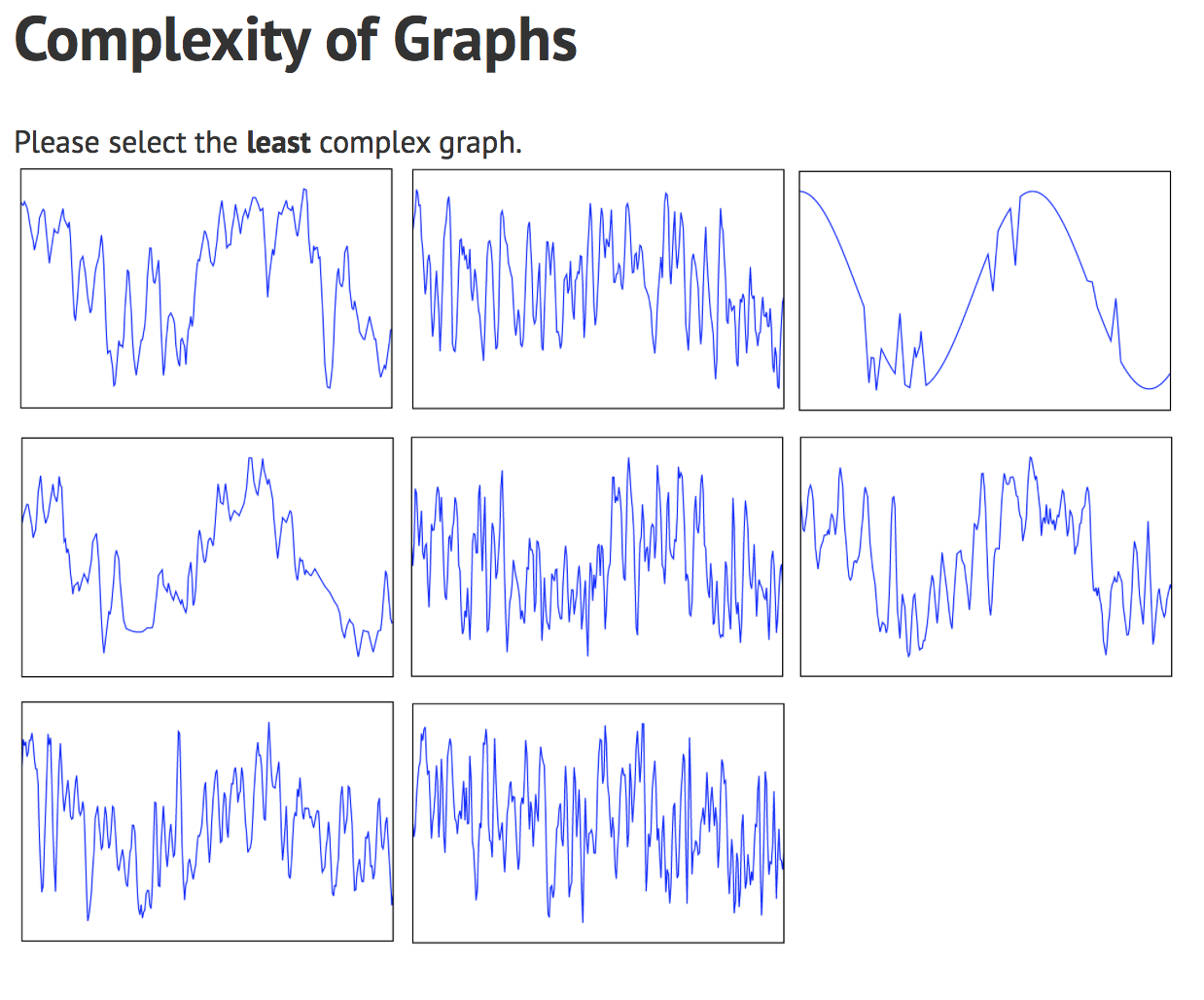}
  \vspace*{-.1in}
  \caption{Example judgement set for cosine function in Experiment 2.}
  \label{fig:graphSet}
 \end{figure}

The synthetic noise tasks generate charts based on five base functions: linear, cosine, gaussian, third order polynomial, and S\&P 500 stock data. For the sample of stock data, we use a 300 sample window of S\&P 500 daily closing prices that has the median PAE taken from a dataset of S\&P 500 stocks over a 15 year period. Each judgment set consists of eight charts from the same base function perturbed with noise to meet a target PAE from 0.1 to 0.8 (in steps of 0.1), and arranged in a random order. For example, the set shown in Figure \ref{fig:graphSet} shows the judgment for the cosine function. 

The real data task used data from three datasets (S\&P 500 historical stock price and volume data, the MIT-BIH Arrhythmia Database, and the Bern-Barcelona EEG Database~\cite{eeg_database,goldberger2000physiobank}), and randomly selected time intervals of the data such that the resulting charts had PAE from 0.1 to 0.8 (in steps of 0.1); they are arranged in random order. 

Each participant was shown the same eight chart judgment set for each baseline function or dataset four times. Two times the participant was asked to select the most complex chart in the set, and two times she was asked to select the least complex. The ordering of judgment sets and position of the target chart was randomized. 
We use the two judgments of the least complex chart as the attention check to prevent Mechanical Turk participants from gaming the system~\cite{gadiraju2015understanding}. If a participant does not select the same chart as least complex during the experiment, we remove the participant's data from consideration.

Participants for the synthetic noise (real data) task made 20 (16) judgments.  50 participants attempted the task; after dropping those that failed the attention check, 33 (25) remained. Of these, 45$\%$ (28$\%$) were female; 76$\%$ (76$\%$) were between the ages of 25 and 49; 55$\%$ (48$\%$) held high school degrees and 36$\%$ ($52\%$) held a Bachelor's degree; and 85$\%$ (80$\%$) spend upwards of 30 hours per week on a computer. Notably, 33$\%$ (52$\%$) ranked themselves as having intermediate expertise in statistical visualizations and 42$\%$ (28$\%$) identified themselves as novices in the field.

\stitle{Results and Statistical Analysis:} 
To test our two hypotheses, we take guidance from~\cite{dixon2008} and perform a binomial logistic regression, where correctness is the outcome variable, and PAE and the baseline function are explanatory variables. Correctness is defined as a binary variable based on whether or not a participant selected the chart with the least PAE as the least complex, and vice versa for most complex.  

The result of the binomial logistic regression (and a follow-up Wald test for baseline function, because it is categorical) indicates a significant association between PAE and correctness $(Z = -4.78, p < 0.001)$, and a significant association between baseline function and correctness $(\chi^2(4) = 10.6, p < 0.05)$ for the synthetic data. Similarly, for the real data there is a significant association between PAE and correctness $(Z = -5.61, p < 0.001)$, but no significant association between base function and correctness.

\stitle{Discussion of Results:} 
We find that PAE might be used to approximate perceived chart complexity. The binomial logistic regression shows that for synthetic and real data PAE is significantly associated with what charts participants select as most or least complex (\textbf{H2.1}). This trend holds for charts from synthetic and real data, and suggests that PAE is a reasonable proxy for perceived complexity of line charts in practice. For synthetic data, we find that underlying base function is significantly associated with what charts participants select as most or least complex (\textbf{H2.2}). 

\begin{figure}[tbh]
  \centering
  \includegraphics[width=\columnwidth]{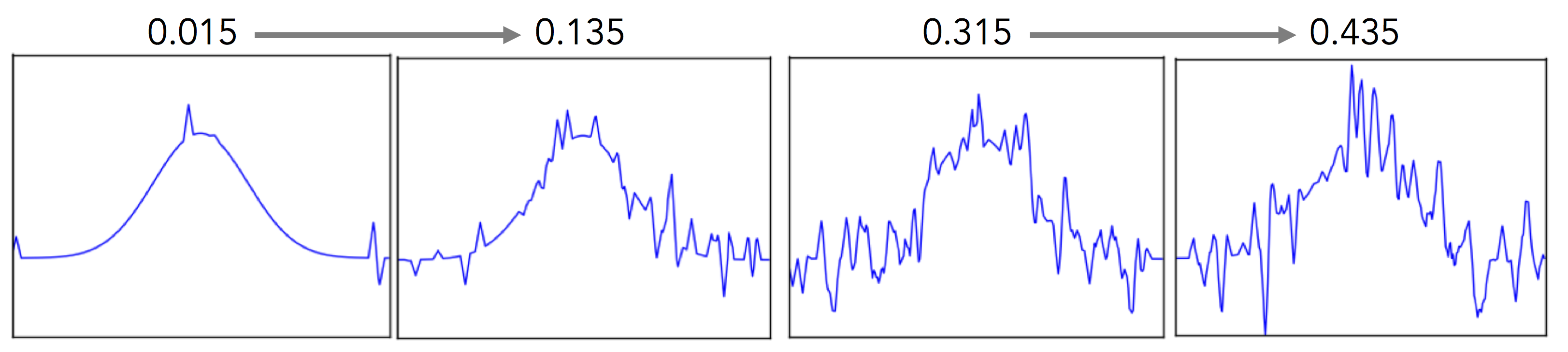}
  \caption{Gaussian function with 0.015, 0.135, 0.315, 0.435 added PAE.}
  \label{fig:gaus}
 \end{figure}

In analyzing results, we notice that participants typically displayed better accuracy in identifying simple charts than complex ones. This suggests that small differences in PAE are easier to spot in charts with lower PAE than with higher PAE. Figure~\ref{fig:gaus} shows this case for a gaussian. Increasing the PAE by $\Delta=0.125$ to a chart with low PAE ($0.015$) is easier to discern than adding $\Delta$ to a higher PAE chart ($0.315$).   This suggests a phenomena akin to just noticeable difference (JND) for PAE, which we explore in later experiments.

%% file: sections/exp3.tex
\begin{figure*}[bt]
  \centering
  \begin{subfigure}[t]{.32\textwidth}
    \centering
    \includegraphics[width=.95\textwidth]{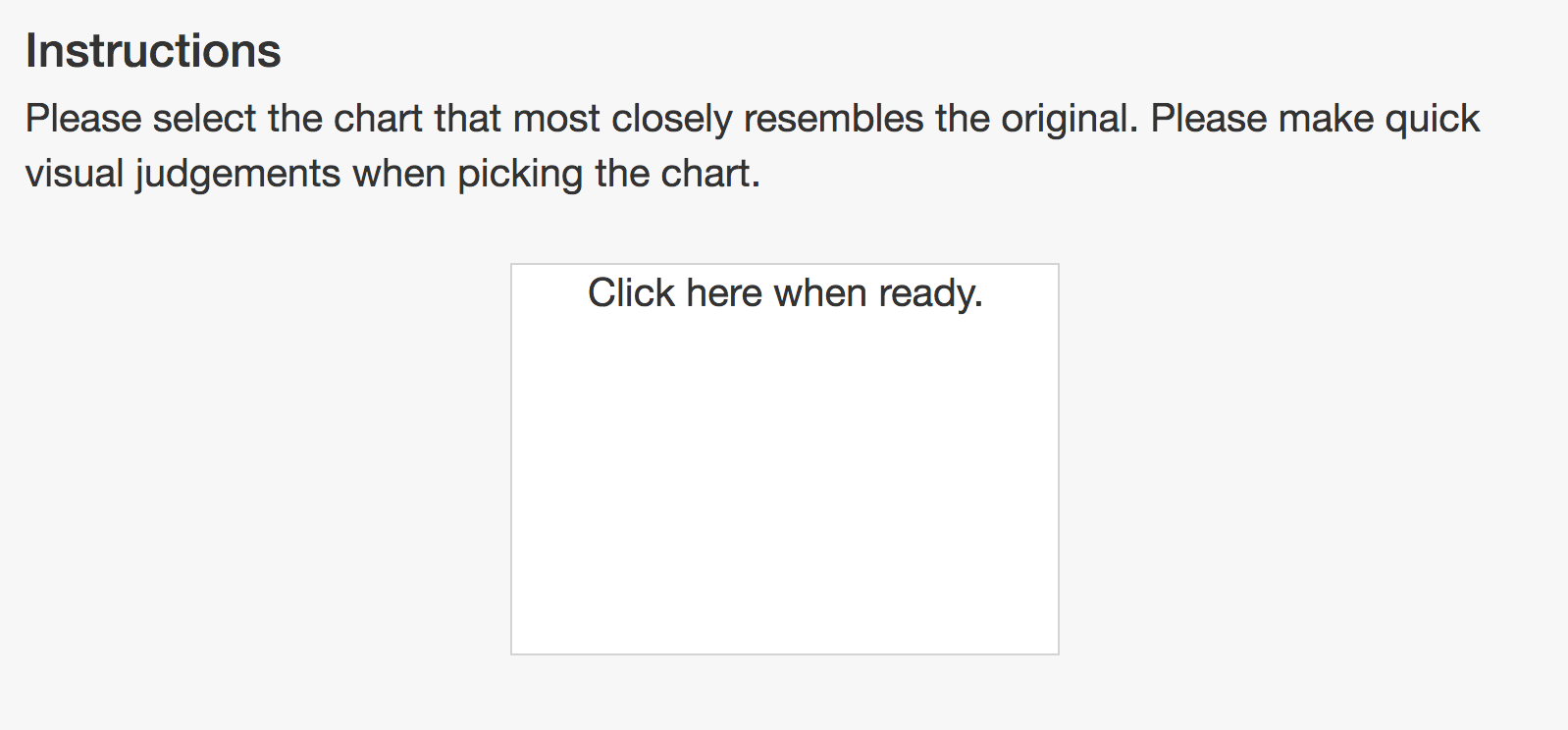}
    \caption{The participant clicks the button to indicate that she is ready to make a judgment.}
    \label{f:exp3_1}
    \vspace*{.1in}
  \end{subfigure}
  \begin{subfigure}[t]{.32\textwidth}
    \centering
    \includegraphics[width=.95\textwidth]{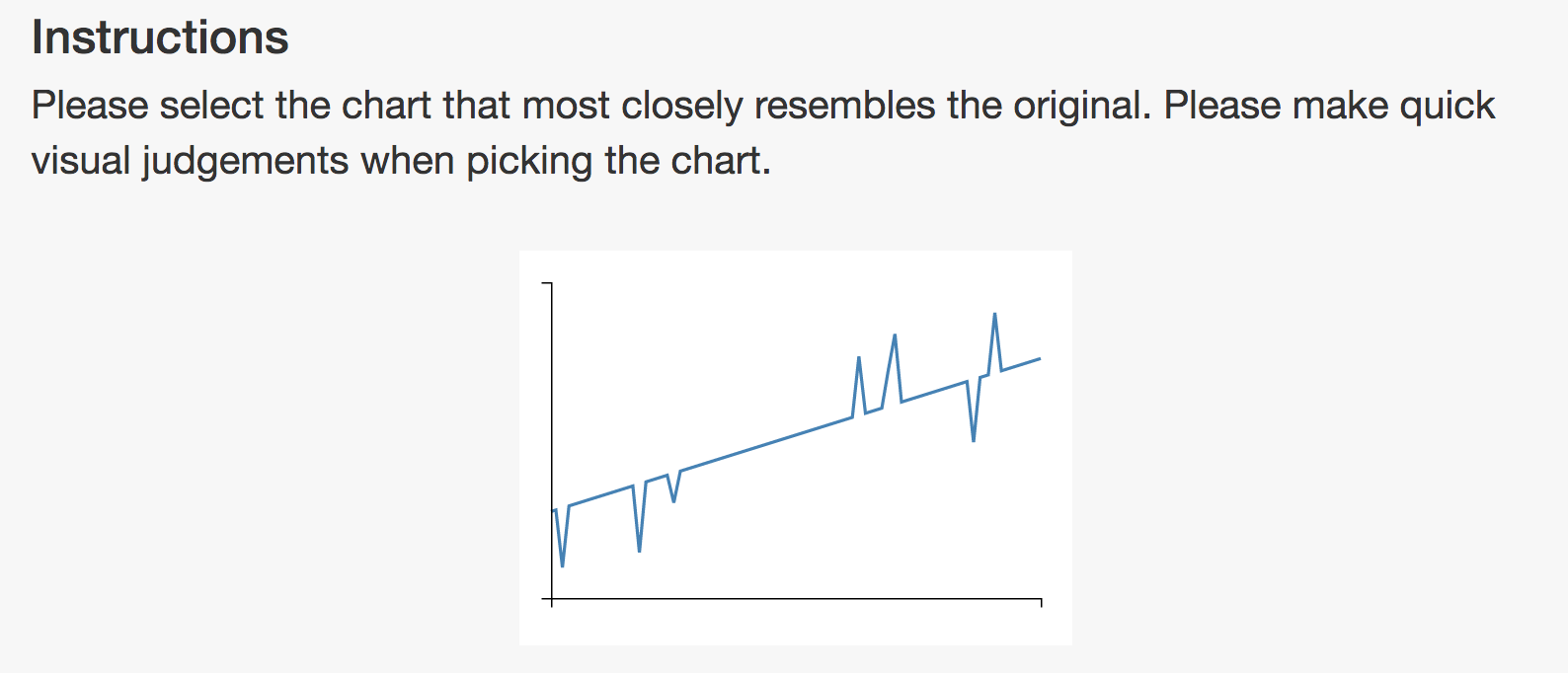}
    \caption{An initial chart is flashed for the {\it glance time}, and then hidden and replaced with a mask during the {\it pause time}.}
    \label{f:exp3_2}
    \vspace*{.1in}
  \end{subfigure}
  \begin{subfigure}[t]{.32\textwidth}
    \centering
    \includegraphics[width=.95\textwidth]{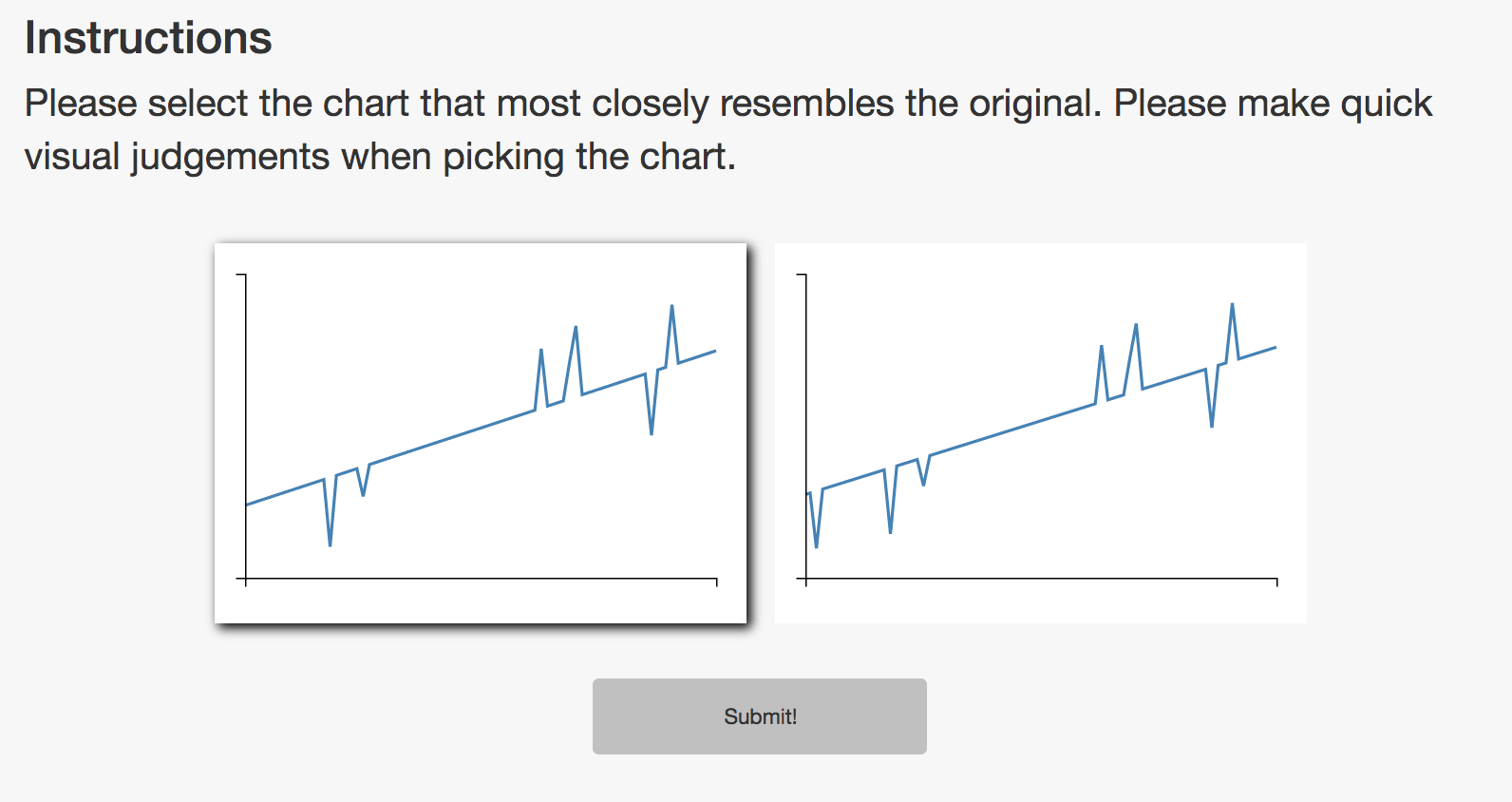}
    \caption{After the {\it pause time}, we show two options: the initial chart and the initial chart with PAE added/removed.  The order of the charts are randomized.}
    \label{f:exp3_3}
  \end{subfigure}
  \vspace*{-.1in}
  \caption{Experiment 3 screenshots. }
  \label{fig:exp3}
\end{figure*}

\subsection{Experiment 3: Find-the-Difference Task}

Do varying levels of chart PAE affect the ability to {\it perform} visual comparison judgements?   In this experiment, we show participants one chart for a short duration (the length of a saccade), hide it for some time using a mask, and then show a copy of the initial chart alongside another chart with higher or lower PAE.   The participant is asked to select the initial chart from the set of two. Participants were given the following direction: ``{\it This HIT consists of a series of graph comparisons. For each comparison, you will first be shown a original chart for a fraction of a second. You then be shown two similar charts and asked to pick the one which most closely resembles the original flashed chart. Please make quick visual judgements and only spend a few seconds when picking the chart.}'' The charts are all generated by adding noise to a base function as specified in the noise generation section. We hypothesized that:

\begin{itemize}[leftmargin=*, topsep=0mm, itemsep=0mm] 
  \item \textbf{(H3.1)} Task accuracy is affected by the PAE of the initial chart
  \item \textbf{(H3.2)} Task accuracy is affected by the magnitude of the PAE difference between the initial and alternative chart.
  \item \textbf{(H3.3)} Task accuracy is affected by the sign of the PAE difference (the alternative chart has higher or lower PAE than the initial)
  \item \textbf{(H3.4)} Task accuracy is not affected by the underlying chart type (linear, cosine, polynomial, gaussian).
\end{itemize}

The setup for Experiment 3 was inspired by Just Noticeable Difference (JND) studies that show two slightly different charts side by side and asking the user to pick the higher chart (for some measure of ``higher'').  By using a staircase protocol that incrementally increases or reduces the differences between the two charts, researchers can find the JND where users are accurate less than 75\% of the time~\cite{rensink2010scatter, harrison2014ranking}. Figure~\ref{fig:exp3} shows screenshots of the experiment.

\begin{figure}[htb]
  \centering
  \includegraphics[width=\columnwidth]{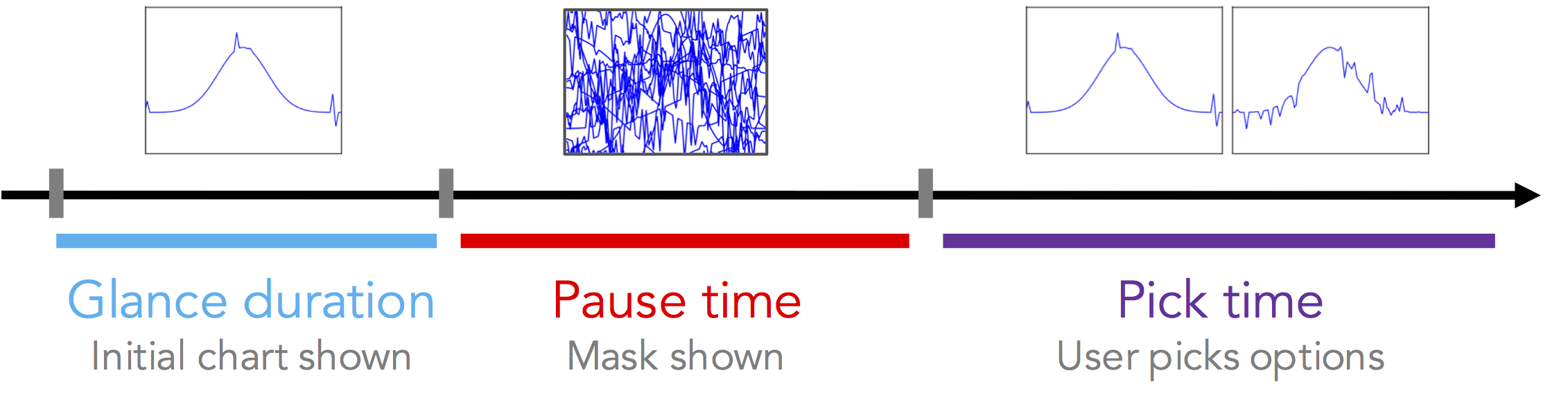}
  \vspace*{-.2in}
  \caption{Timeline of the at a glance task.  We first show the initial chart for {\it glance time}, then hide the chart and don't show anything for a {\it pause time}, and then show the participant options to choose from.}
  \label{fig:timeline}
 \end{figure}

Each judgment follows the process illustrated in Figure~\ref{fig:timeline}. Participants first view an initial chart that flashes for a glance time of 200ms, the duration of one saccade, or slightly longer than the basic stage of correlation perception of 150ms proposed by Rensink et al., and the amount of time required for eye movements to be optimially guided~\cite{rensink2016entropy, rayner2009}. The initial chart is replaced with a mask that is shown for 200ms of pause time.  Masking is used to interrupt the perceptual processing so that user responses are due to cognitive, rather than low-level perceptual pattern matching~\cite{renninger2004scene,schyns1994blobs,gheorghiu2008size,broers2018enhanced,potter1975meaning,potter1976short}.  The mask, shown in Figure~\ref{fig:timeline}, is designed to be similar in visual structure as a line chart, but does not convey any information related to the user task.   After the pause time, the participant is shown the initial chart and the initial chart with $\Delta_y$ more or less PAE, and asked to choose the initial chart.  The order of the two charts is randomized to reduce learning effects.

Our protocol differs from classic JND protocols in that it focuses on the ability to identify a change in chart complexity after the initial glance time.
Further, since our goal was not to find this just noticable threshold (although that is a direction for followup work), we did not directly follow the staircase protocol.  We instead sweep through different chart PAE values to confirm that task accuracy is affected by chart PAE.

Based on pilots, the initial chart has low ($0.045$), medium ($0.09$), and high ($0.18$) PAE.  The comparison chart differed in PAE by $\Delta_y$ of $\pm0.015$, $\pm0.03$, $\pm0.06$, $\pm0.09$, and $\pm0.12$. We evaluate these conditions for all four chart types (line, cosine, gaussian, and third order polynomial), and both increasing and decreasing PAE, resulting in a $3\times 5\times 2\times 4=120$ factorial design.

\stitle{Results and Statistical Analysis:}
There were $52$  participants, who took on average 7 minutes and 5 seconds to complete the tasks in the experiment, or approximately 3.5 seconds to complete each comparison task. The participants had the following demographics: 63\% male; 65\% between 25 and 39 years old; 58\% held Bachelor's or Master's degrees; and per-week computer usage was fairly even from $21$ to $>60$ hrs. Additionally, 67\% ranked themselves as low-intermediate, intermediate, or high intermediate level visualization users.

In data collection we recorded the correctness of a participant's response for each judgment, and used this as the binary response variable in our analysis. To test our four hypotheses we performed a binomial logistic regression where we used the base function, initial chart PAE, magnitude of PAE difference, and the sign of the difference, to predict task accuracy. We also used a a follow-up Wald test for baseline function, because it is categorical.  We found that all independent variables are predictive of task accuracy.  Their correlation and significant scores are summarized as:
\textbf{(H3.1)} Entropy of the initial chart $(Z = 2.42, p < 0.05)$; 
\textbf{(H3.2)} Magnitude of PAE difference between the two charts $(Z = -5.08, p < 0.001)$; 
\textbf{(H3.3)} Sign of the PAE difference between the two charts $(Z = -8.06, p < 0.001)$;  and
\textbf{(H3.4)} Baseline function $(\chi^2(3) = 9.32, p < 0.05)$. 

\begin{figure}[h]
  \centering
  \includegraphics[width=.9\columnwidth]{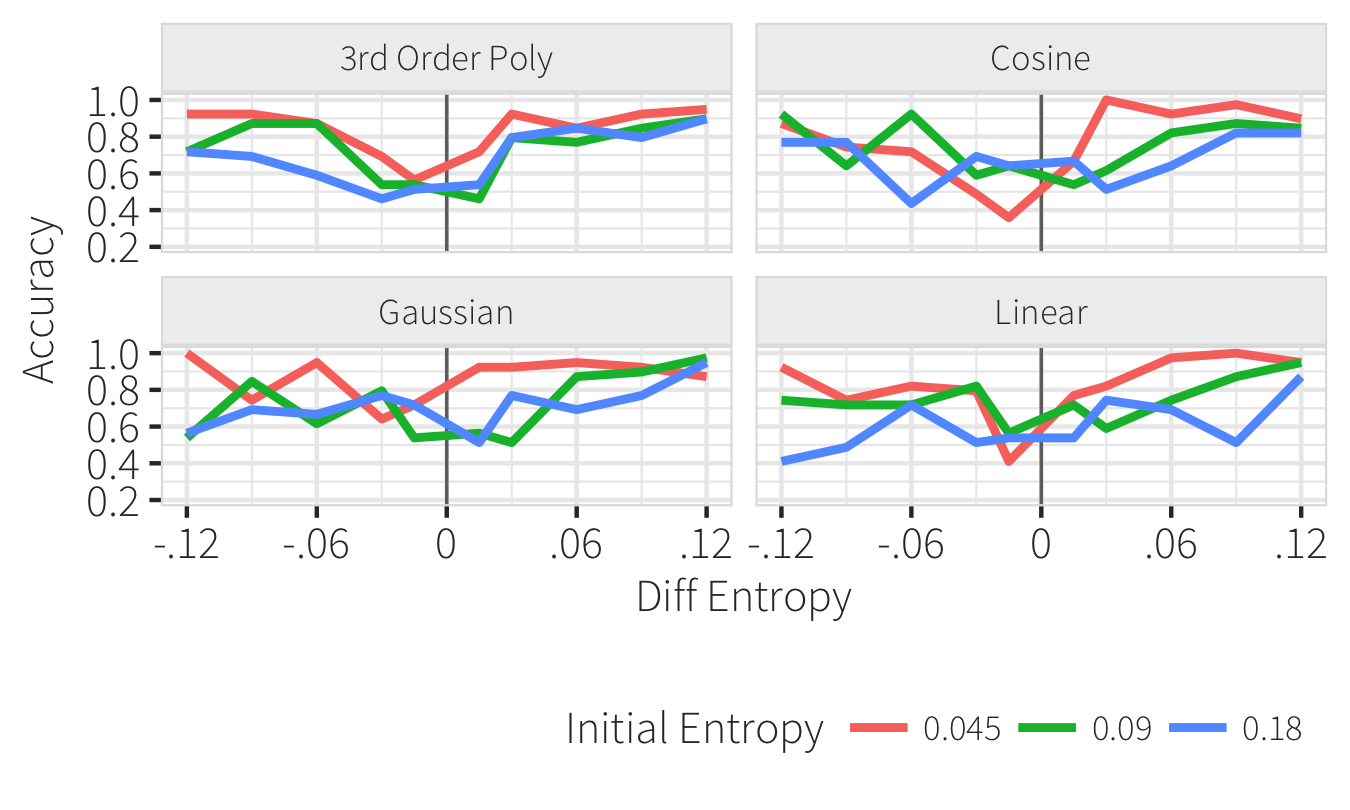}
  \vspace*{-.2in}
  \caption{Average accuracy as $\Delta_y$ of PAE varies.  Each facet shows a different chart type, and each line is a different initial PAE.}
  \label{fig:exp3_entropy}
\end{figure}

These results imply that the initial chart's PAE, magnitude and sign of $\Delta_y$, and chart type have an effect on participant's accuracy in identifying the initial chart in the gallery. Figure~\ref{fig:exp3_entropy} illustrates a clear effect of initial PAE on accuracy. Higher initial PAE (blue) systematically reduces accuracy as compared to low initial PAE (red), especially for the linear chart when $\Delta_y$ (x-axis) is small.  Accuracy also increases as the magnitude $|\Delta_y|$ increases, meaning it is easier to identify large changes in visual complexity.  Across all conditions, we find that judgment accuracy converges to $0.5$ (random chance) as $\Delta_y$ decreases towards $0$.

\begin{figure}[htb]
\centering
\includegraphics[width=.9\columnwidth]{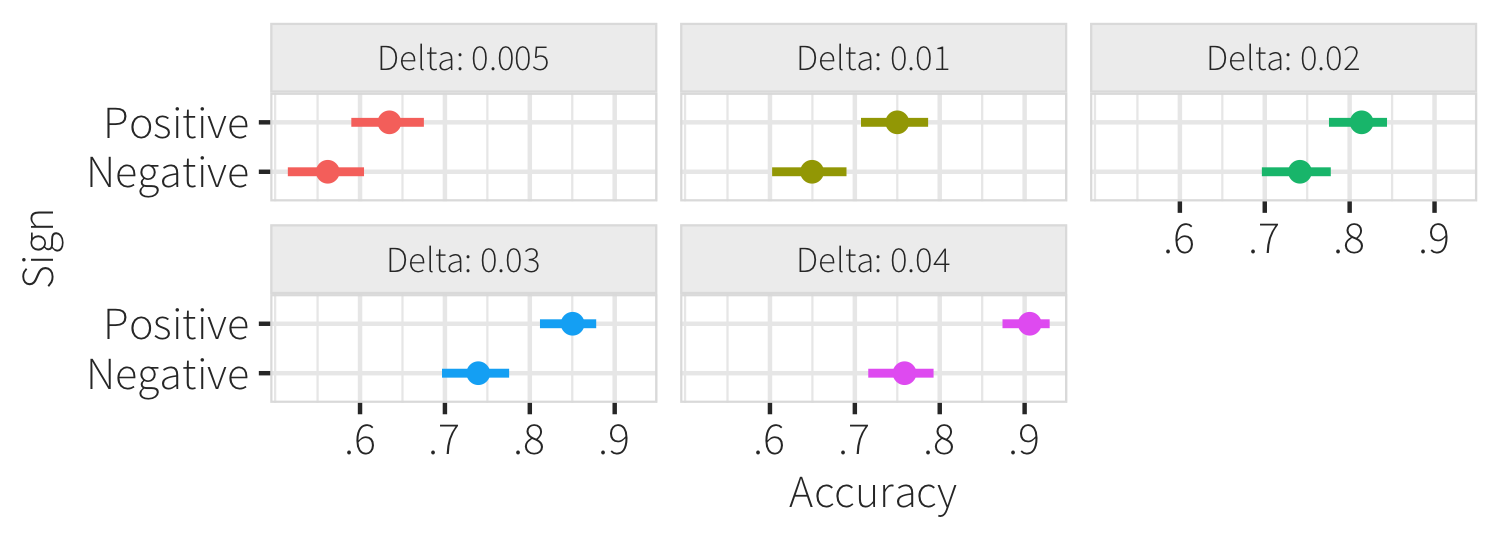}
\vspace*{-.2in}
\caption{Mean and $95\%$ bootstrapped confidence interval of judgement accuracy for increasing (positive) or reducing (negative) the PAE of the alternative chart options. Each facet is a different $\Delta_y$ magnitude.}
\label{fig:exp3_sign}
\end{figure}

Interestingly, the sign of $\Delta_y$ affects task accuracy.  To investigate this, we grouped participant judgments by whether $\Delta_y$ is positive (the alternative chart has higher PAE) or negative (vice versa).  Figure~\ref{fig:exp3_sign} plots the $95\%$ bootstrapped confidence intervals, faceted by the magnitude of the PAE difference $|\Delta_y|$. 
When $|\Delta_y|$ is low ($\le0.01$), participants are more accurate when the change is positive than negative.  This happens because participants preferentially choose the lower PAE chart, which is correct when $\Delta_y$ is positive, and incorrect when negative. 
As the magnitude of the PAE difference increases, the accuracy for both signs increas, however the bias also persists.

\stitle{Discussion of Results:}
Based on the results, we accept all four hypotheses: \textbf{H3.1}, \textbf{H3.2}, \textbf{H3.3}, and \textbf{H3.4}.  
We find a clear trend towards 50\% accuracy (random guess) as the difference in PAE between the two charts decreases, irrespective of other conditions.  
Additionally, as the initial PAE increases (the chart is more complex), so too does the necessary change in PAE in order to accurately differentiate the charts.   This supports our finding in Experiment H2, that participants are worse at distinguishing between changes for high PAE charts than they are for low PAE charts or when the difference in PAE is large.

Further, given two charts, participants have a systematic bias towards choosing the chart with lower PAE. We do not have an explanation of why this might be, but hypothesize that it might be because the lower PAE charts appear closer to the ``shape envelope'' of the initial chart, especially when viewed at a glance.   However, further studies are needed to evaluate this conjecture.

%% file: sections/exp4.tex
\subsection{Experiment 4: Shape Identification Task}

This experiment uses the same procedure as experiment 3, but for a different visual judgment task: identifying the overall shape of a chart. Many applications of visualizations involve users attempting to identify patterns in potentially noisy data, and we designed this experiment to simulate this use case. To this end, we provided the participants with examples of 4 underlying base functions (shapes) to study. Then, after adding enough noise to a base function to reach a desired PAE level, users are asked to identify which of the 4 underlying functions the noisy chart represents. The hypothesis is that the PAE of the chart affects the ability to accurately identify the underlying function of the chart (\textbf{H4}). Note that in the extreme, there can be so much noise that the charts are quantitatively identical irrespective of the initial base function, and the accuracy should converge to random guessing.

\begin{figure}[t]
  \centering
  \begin{subfigure}[t]{\columnwidth}
    \centering
    \includegraphics[width=\columnwidth]{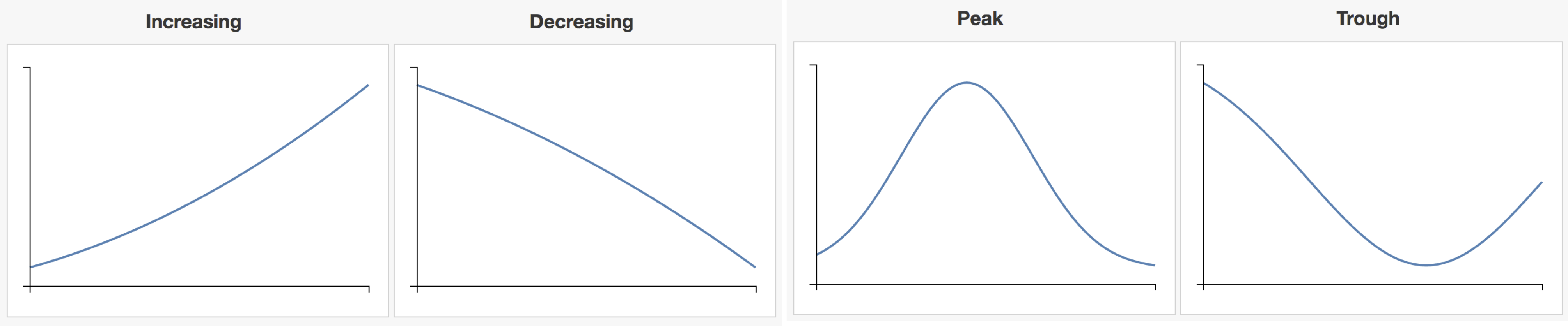}
    \caption{The four basic chart shapes users needed to identify.}
    \label{f:exp4_baselines}
    \vspace*{.1in}
  \end{subfigure}\\
  \begin{subfigure}[t]{.45\columnwidth}
    \centering
    \includegraphics[width=.85\columnwidth]{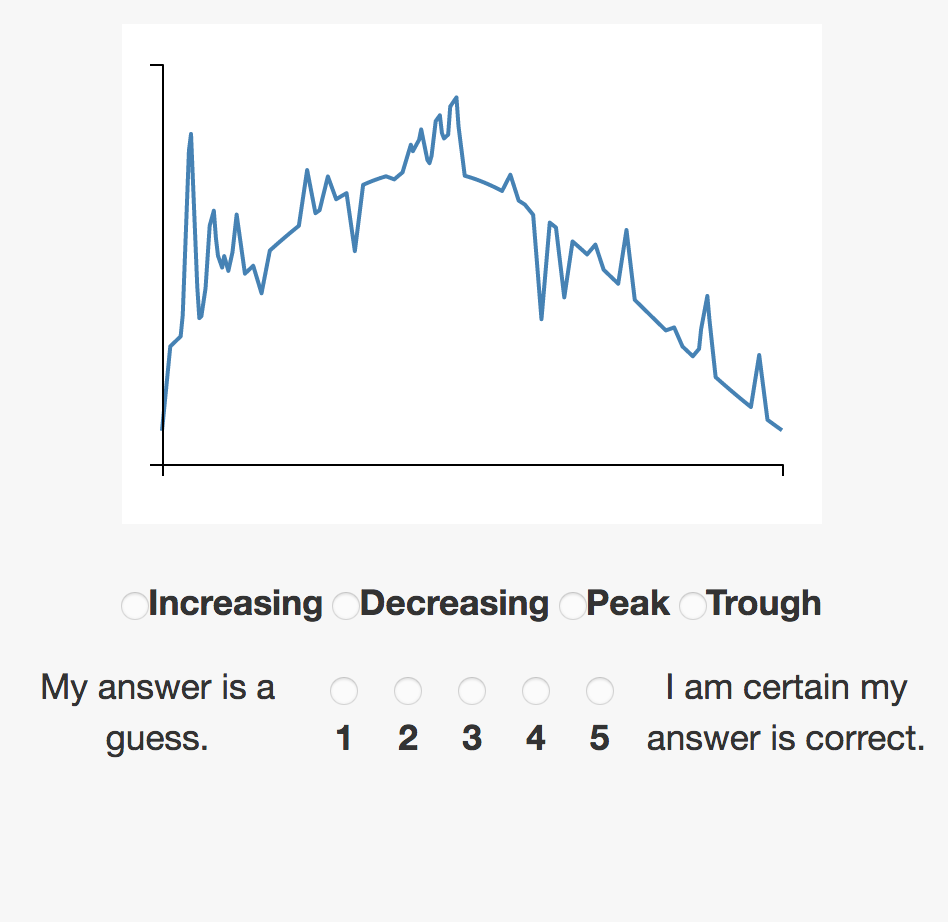}
    \caption{Peak shaped chart with 0.2 PAE.}
    \label{f:exp4_1}
    \vspace*{.1in}
  \end{subfigure}
  \begin{subfigure}[t]{.45\columnwidth}
    \centering
    \includegraphics[width=.85\columnwidth]{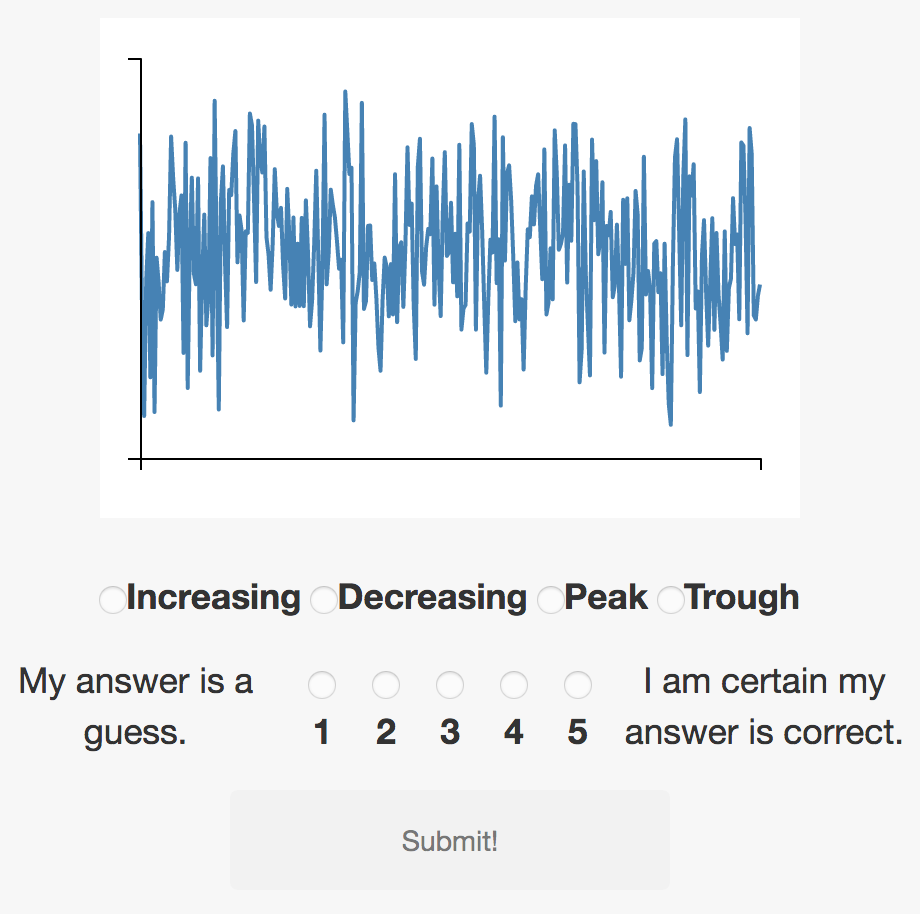}
    \caption{Increasing trend chart with 1.2 PAE.}
    \label{f:exp4_2}
  \end{subfigure}
  \caption{Screenshots depicting Experiment 4. }
  \label{fig:exp4}
\end{figure}

\stitle{Participant Tasks:} 
Participants first view an initial chart that flashes for a {\it glance time} of $500$ ms ($\approx2-4$ fixations). We allowed a slightly longer glance time since identifying the underlying shape of a chart is more difficult than the previous task of matching charts. The participant is then asked which of the following four shapes is most representative of the chart: increasing trend, decreasing trend, peak, and trough (Figure~\ref{fig:exp4}).  They are also asked to self-report their answer certainty, from $0$ for `guess' to $4$ for `certain'.  Chart order was randomized to reduce learning effects. Participants were asked to make quick judgments, and we limited their time to complete the experiment 30 minutes maximum. For each base function, we showed users $5$ versions at each of four PAE levels ($0.2$, $0.4$, $0.8$, $1.2$), resulting in a $4\times 4\times 5=80$ factorial design.

\stitle{Results and Analysis:} 
There were $47$ total participants, who took on average 8 minutes to complete the tasks in the experiment, and approximately 3.6 seconds to complete each comparison task. The participants had the following demographics: 62\% fell between the ages of 25 and 39, computer usage was spread evenly, although 88\% used computers for at least 20 hours per week, 57\% held Bachelor's or more advanced degrees, 76\% were male, and 79\% ranked themselves as some level of intermediate user with statistical visualizations.

\begin{figure}[tbh]
  \centering
  \includegraphics[width=\columnwidth]{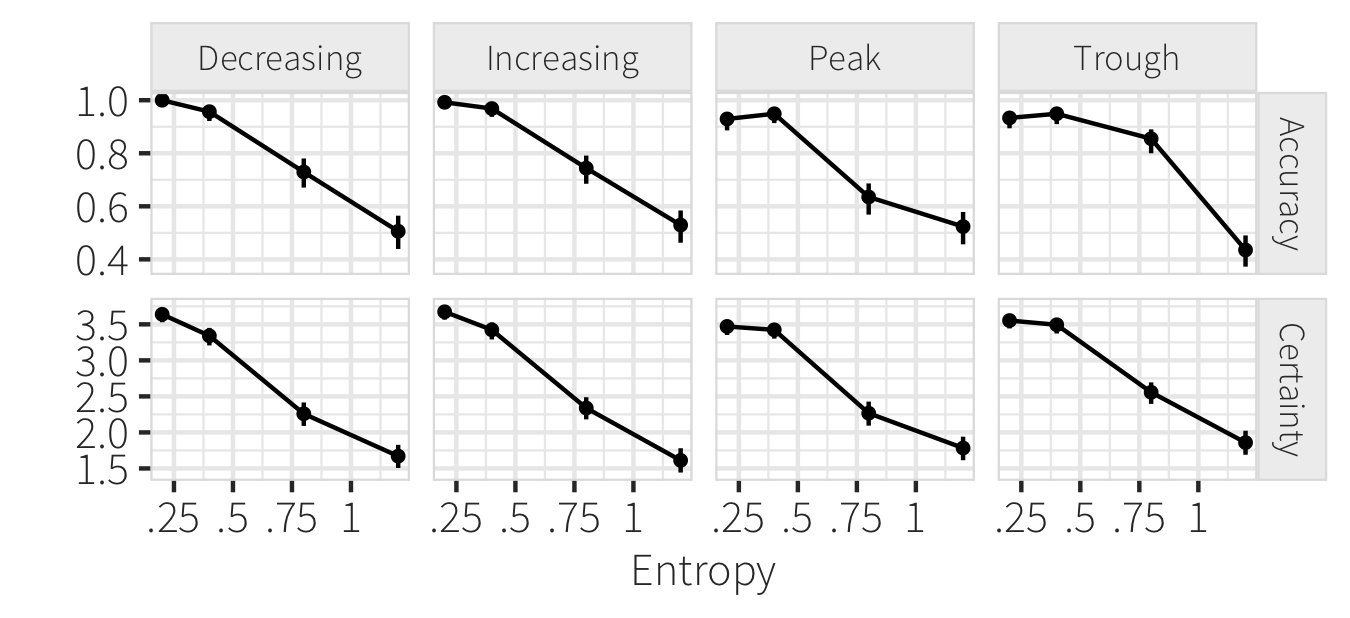}
  \vspace*{-.2in}
  \caption{Chart shape identification accuracy and certainty as entropy changes.  Each facet shows a different chart type.}
  \label{fig:exp4_accuracy}
\end{figure}

We performed a binomial logistic regression using PAE as the independent variable to predict task correctness as the dependent variable. We found a significant correlation $(Z = 25.75, p < 0.001)$, which supports \textbf{H4}.  
Figure~\ref{fig:exp4_accuracy} shows that users were able to correctly identify the underlying shape of a chart with close to $100\%$ accuracy when the PAE (x-axis) of the chart, with the added noise, is low. However, user accuracy drops to $\approx70\%$ for a PAE of 0.8, and to $\approx50\%$ for a PAE of 1.2.    Further, we find that answer certainty consistently decreases as the PAE increases.

\stitle{Discussion:} 
For the shape identification task, we found that there is a clear trend towards lower accuracy in distinguishing chart shapes as the entropy increases, indicating that users find it more difficult to perceive meaningful differences in charts with high entropy.  
We expect that there is a maximum PAE where charts are quantitatively the same irrespective of the base function, and users resort to random guessing.  We do not believe our experiment approached this limit, because the lowest accuracy was still higher than random chance (25\%), and the lowest average certainty was $\approx1.5$ rather than $0$. Looking further, this study primarily added high frequency random noise to base functions that are low frequency.  We speculate that adding lower frequency noise, such as alternative base functions, may reduce identification accuracy at PAE levels.

%% file: sections/exp5.tex
\subsection{Experiment 5: Glance and Pause Time}

We now turn to the interaction between glance (and pause) time with PAE on the user's ability to perform visual judgments.   We hypothesize that the ability to perceive differences in charts is affected by \textbf{(H5)} the length of glance time. 
We use the same task as Experiment 3, but vary the initial chart's glance time.
We ran a separate user study that varied the pause time between the inital chart and the two chart options; we found that there was no discernable difference, nor any statistically significant effect, due to the pause time.  We thus omit the details due to space constraints.  

\stitle{Participant Tasks:}
We fixed the baseline function type to linear; varied initial PAE levels $\in$ \{$0.045$, $0.09$, $0.18$\}; varied $\Delta_y \in$ \{$0.015$, $0.06$, $0.24$\}; and showed the two chart options immediately after the mask is hidden (pause time is $0$). We varied glance time $\in$ \{$50ms$, $100ms$, $200ms$, $2s$\} to allow for different numbers of saccades: no more than one fixation to enough time to study the visualization. The study was a $3\times3\times2\times4=72$ factorial design.

\begin{figure}[htb]
  \centering
  \includegraphics[width=\columnwidth]{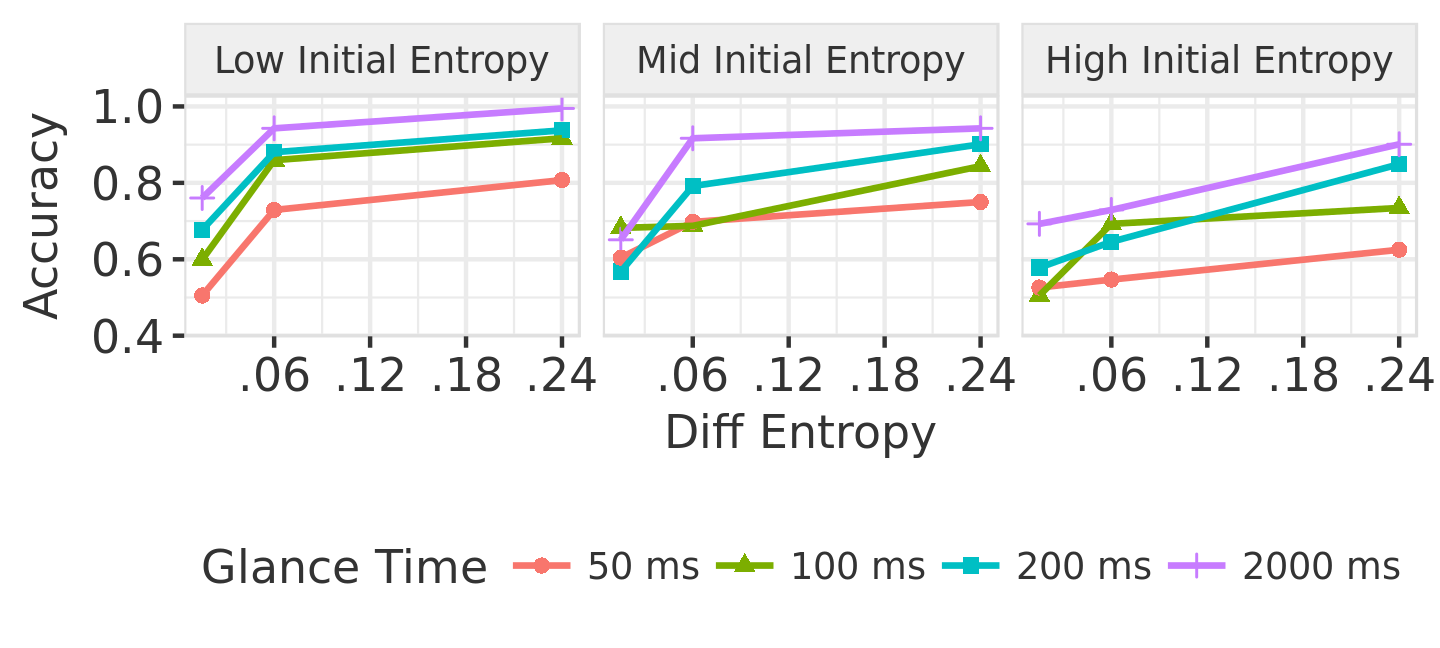}
  \vspace{-.4in}
  \caption{Acc vs $\Delta_y$ for glance time (lines) and initial PAE (facets).}
  \label{fig:exp5_duration}
 \end{figure}

\stitle{Results, Analysis, and Discussion:}
There were 48 participants, who took on average 8 minutes and 21 seconds to complete the tasks in the experiment, or approximately 3.5 seconds to complete each comparison task. There were 81\% between the ages of 25 and 39, and 56\% classifying themselves as intermediate visualization users. We performed a binomial logistic regression, with glance time as the independent variable to predict response accuracy.  Glance time $(Z = -4.01, p < 0.001)$ was significantly correlated with accuracy and supports \textbf{H5}.
Figure~\ref{fig:exp5_duration} shows that as the initial PAE level increases (left to right facets), glance time (line) affects judgement accuracy more.  As before, increasing $|\Delta_y|$ makes the judgement easier, and increases accuracy.  
We find that glance time has a significant effect on the accuracy of identifying differences in the initial chart. At short glance times ($20$ms), users are nearly unable to discern small changes in PAE or when the initial PAE is high. Although longer glance times show higher accuracy, the task is still challenging when the initial PAE is high.  
We speculate that further increasing the glance time (e.g., to seconds or minutes) may improve accuracy further, but there may be a limit to the accuracy when the initial entropy is high.

%% file: sections/discussions.tex
\section{Discussion}

Our findings on PAE and the perception of line chart complexity have implications for visualization design.  They help bridge the gap between research on low-level perception and high-level visualization, and provide a user (as opposed to data) centered measure of visualization complexity. We describe several possible applications for a quantifiable visual complexity measure: 

\stitle{Highlighting Changes:} PAE helps quantify changes in visualization complexity, and may be used when it is important to ensure that the user understands changes and comparisons.    If the visualization is too complex to perceive certain changes, they may need to be emphasized to the user. Similarly, the systemic bias users experience in differentiating charts of increasing or decreasing complexity implies that visualizations may want to more emphasize changes that increase, rather than decrease, complexity. Further, if PAE is high, users may have difficulty identifying the underlying function in the visualization---if such a detail is important, designers should specifically draw attention to such differences.      
Finally, the glance time findings can inform the design of displays that rapidly show or change charts.  Users need more time to understand more complex charts, and PAE can inform designers to be aware of the user's exposure time depending on chart complexity, and perhaps highlight the key changes.

\stitle{Large Dataset Visualization:} The results can also apply to interactive and approximate visualizations of very large data sets. Quantifying the level of complexity at which a user can identify changes can help designers of approximate visualizations~\cite{ding2016sample,hellerstein1997online,procopioload,alabi2016pfunk} determine when further sampling or other computation will no longer yield perceptible differences.  Similarly, designers of interactive visualizations might develop optimizations based on this same phenomena.  

\stitle{Visualization Parameter Selection:} PAE might help guide visualization parameter selection (e.g., aspect ratios, layering). For example, the choice of horizon graph~\cite{heer2009sizing} height and layering may be informed by measuring the resulting chart complexity using PAE.   Likewise, PAE could inform aspect ratio selection methods to select a ratio that takes visual complexity into account.  

\stitle{Guided Smoothing:} Guided Smoothing: Finally, PAE can be used to indicate when, and to what extent, smoothing or other simplifying methods may be applicable to given chart. For example, PAE could be used with a smoothing method like ASAP to determine when ASAP should be applied to make a chart easier to read, or alternately, PAE could be incorporated into an iterative smoothing method to detect when the chart has simplified enough to be easily readable ~\cite{rong2017asap}.

%% file: sections/conclusions.tex
\section{Conclusions and Future Work}

This paper studied the perception of complexity in line chart visualizations. We derive a new measure for visual complexity based on approximate entropy, Pixel Approximate Entropy.  We conduct analytical and user experiments to validate its suitability as a complexity measure. In particular, we look at using PAE as a measure of visual complexity; users' ability to perceive minute differences in complexity; and the effect of time on a users' ability to perceive differences in complexity. We performed four sets of experiments and found that PAE correlates closely with noise, that as PAE of a chart increases so too does perceived complexity of the chart; users are better able to perform visual comparison tasks when a chart's PAE is low than when it is high; and users have more difficulty with visual comparison on charts with high PAE when they have less time to view a chart.

There are a number of ways we intend to extend our findings. The first is to design and perform a formal stair case study to better understand whether the JND limits for PAE follow Weber's law. Staircase studies have been used to identify other perceptual features that follow Weber's law, for example, Harrison et al. used a large scale staircase study to measure the JND limits for perception of correlation across 9 visualization types~\cite{harrison2014ranking}. We intend to perform a similar study to investigate perception of complexity and its relationship with PAE across different one dimensional visualzations. PAE could also be applied to predicting user's perception of correlation in line charts. Rensink et al's studies on scatter plots suggests that users in fact rely on perceived entropy to estimate correlation, whether PAE has similar relationship in 1D visualizations is worth investigating~\cite{rensink2017}. 

The second possible extension for our work is to modify PAE to work with other marks and visual encodings. We chose to study line charts in this work because the pixel values rendered in a line chart naturally mapped to a vector of pixel values that could be used to measure PAE.  However, line charts are simply one of many possible ways to visually encode data and directly quantifying the visual complexity of other types of visualizations, such as bar charts, scatter plots, and pie charts, remains an open problem.

We also intend to investigate the potential applications of PAE to designing more effective visualizations. As noted in the discussion, PAE has potential applications in designing charts that communicate changes more effectively, efficiently visualizing large data sets, and guiding visual parameter selection and smoothing or other simplifying operations. Testing these applications will require developing visualization systems that integrate PAE and conducting extensive user studies. Ultimately, by developing a quantifiable measure of visual complexity, we hope to contribute to systems that can guide visualization designers in making more readily understandable charts and automatically generate readable charts on their own.